\documentclass[10pt, a4paper, preprint]{article}
\usepackage[utf8]{inputenc}
\usepackage{multirow}
\usepackage[T1]{fontenc}
\usepackage{threeparttable} 
\usepackage{graphicx}
\usepackage{authblk}
\usepackage{natbib}
\usepackage{caption}
\usepackage{subcaption}

\title{Omnidirectional Radiation Detector with Perpendicular Dual Silicon Photomultiplier Readout - Directional Sensitivity and Machine Learning Source Positioning}

\author[1,*]{Ana Marija Kožuljević}
\author[1]{Gabriela Jazvac}
\author[1]{Luka Lotina}
\author[1,2]{Luka Pavelić}

\affil[1]{Institute for Medical Research and Occupational Health, Division of Radiation Protection\\Ksaverska c. 2, 10 000 Zagreb, Croatia}
\affil[2]{Nucleometrix d.o.o.\\ Papova 4, 10 000 Zagreb, Croatia}
\affil[*]{Corresponding author: akozuljevic@imi.hr}

\begin{document}
\maketitle

\begin{abstract}
Directional information of incident gamma rays in radiation detection is essential to many applications, from medical imaging to mapping radioactive contamination after nuclear accidents, characterization of nuclear waste during decommissioning, and nuclear security. In this work, we present a radiation detector capable of detecting gamma-ray photons in 4$\pi$, utilizing Gadolinium Aluminum Gallium Garnet (GAGG) scintillating crystals and silicon photomultipliers (SiPMs). The GAGG crystals are assembled in a cubical 4$\times$4$\times$4 matrix, and their light output is read out by SiPMs from two neighboring sides. The size of each crystal is 3$\times$3$\times$3 mm$^3$, while the matrix pitch is 3.2 mm, matching the size and the pitch of the 4$\times$4 SiPMs for one-to-one coupling. The layers perpendicular to the SiPMs are separated by optical reflectors, providing efficient light collection while lowering the probability of inter-crystal leakage. The design of the radiation detector offers high detection efficiency and full-view imaging of gamma-ray sources through Compton scattering. Monte Carlo simulations were performed in Geant4 to evaluate the detection efficiency of the proposed design due to its geometrical non-uniformity. Machine learning models developed with the XGBoost algorithm were trained and tested on the simulated data to assess the capability of the detector to localize point sources of 511, 662, and 1275 keV energies. We find that the proposed design does not influence the collection efficiency of the high-energy gamma-ray photons and offers sensitivity to the direction of the incoming gamma rays. The trained models are capable of determining the source-to-detector distance of the Na-22 and Cs-137 point sources from measurements, thus providing initial conditions for faster image reconstruction.

\end{abstract}

\section{Introduction}

Knowledge of directional information is a crucial element in the imaging of gamma-emitting sources, occurring in many applications of radiation detection, from medical imaging and nuclear security to nuclear decommissioning and post-nuclear disaster decontamination of released radionuclides, such as Cs-137 \cite{uher,sato}. Common methods of detection include position-sensitive, energy-resolving detectors, often limited in their field of view (FOV) and detection sensitivity, requiring heavy collimators or coded aperture approaches. These methods need multiple detectors or additional equipment for rotation to cover the full FOV, creating further complications such as increased weight and price, as well as longer acquisition times. In this work, we present a novel type of directional 4$\pi$ radiation detector, covering the full FOV with depth of interaction (DOI) information, without additional appliances (Fig. 1). The design relies on detecting the scattered and absorbed gamma rays in one volume of a segmented scintillating detector, read out from two sides by silicon photomultipliers (SiPMs). It consists of 64 voxels, each voxel being a Gadolinium Aluminum Gallium Garnet (GAGG) cube. The SiPMs are perpendicular to each other, sitting on two neighboring faces of the cubical volume. Other similar cubical designs have more complicated readout configurations, for example, the SiPMs coupled to the four lateral faces of the detector \cite{peng}, or all six of the faces covered by SiPMs \cite{yamaya, ghelman, GHELMAN2025170246}. 

The presented radiation detector has the modality of a Compton Camera (CC), a gamma-ray imaging device that utilizes electronic collimation. Typically, a CC has two layers: a layer in which the incoming gamma ray scatters and transmits part of its energy to the recoil electron in the material, and a second layer in which the gamma ray is photoelectrically absorbed. Each layer in this configuration has its own readout, but this design is still limited to a small FOV and low detection efficiency, as the sensitive component of the CC is only one side of the detector exposed to incoming gamma rays. As each voxel cube of the proposed detector design can act as a scatterer and an absorber, the detection efficiency is enhanced, and the compactness of the detection system is improved. Unlike other proposed parallel dual read-out CCs \cite{kishimoto, lee3}, this design does not require calculating signal ratios for DOI determination. The capability of this radiation detector design to act as a CC has been previously explored, and images of a Cs-137 point source positioned in front of one face and one vertex have been reconstructed \cite{hndpaper}.
To reconstruct an image of a radiation source, reconstruction algorithms, such as simple and forward back-projection, require information of the source position before the reconstruction \cite{RAWOOLSULLIVAN20081986, DASH2023167808}. Maximum likelihood estimation methods (like Maximum Likelihood Expectation Maximization - MLEM) do not require prior knowledge of the source position, but providing initial values would greatly reduce computing requirements and reconstruction time. In this paper, we utilize the eXtreme Gradient Boosting (XGBoost) machine learning library for determination of the source-to-detector distance from raw measurements. We show that the detector itself is sensitive to the direction of the incoming gamma rays through attenuation in its sensitive volume, enabling source positioning prior to the reconstruction. Machine learning (ML) methods have been successfully applied to problems of localization and identification of gamma-ray sources \cite{xai,ml-nearfield1, ml-nearfield, DURBIN2022166905}, as well as for event selection and detection optimization \cite{llosa,hpge_acc}. XGBoost library has been previously utilized for tracking of the interactions in scintillator-SiPM particle detectors \cite{sspd-ml}, but its capability for source-to-detector distance determination is explored here for the first time. We also examine the detection efficiency of the proposed design for gamma-ray energies of Na-22 and Cs-137 point sources. The efficiencies at these energies are relevant for experimental measurements, as two sides of the sensitive volume are covered by SiPMs. 

\section{Methods}
\label{sec:methods}

\subsection{Detector design}

\begin{figure}[t]
  \centering
  \includegraphics[width=0.7\linewidth]{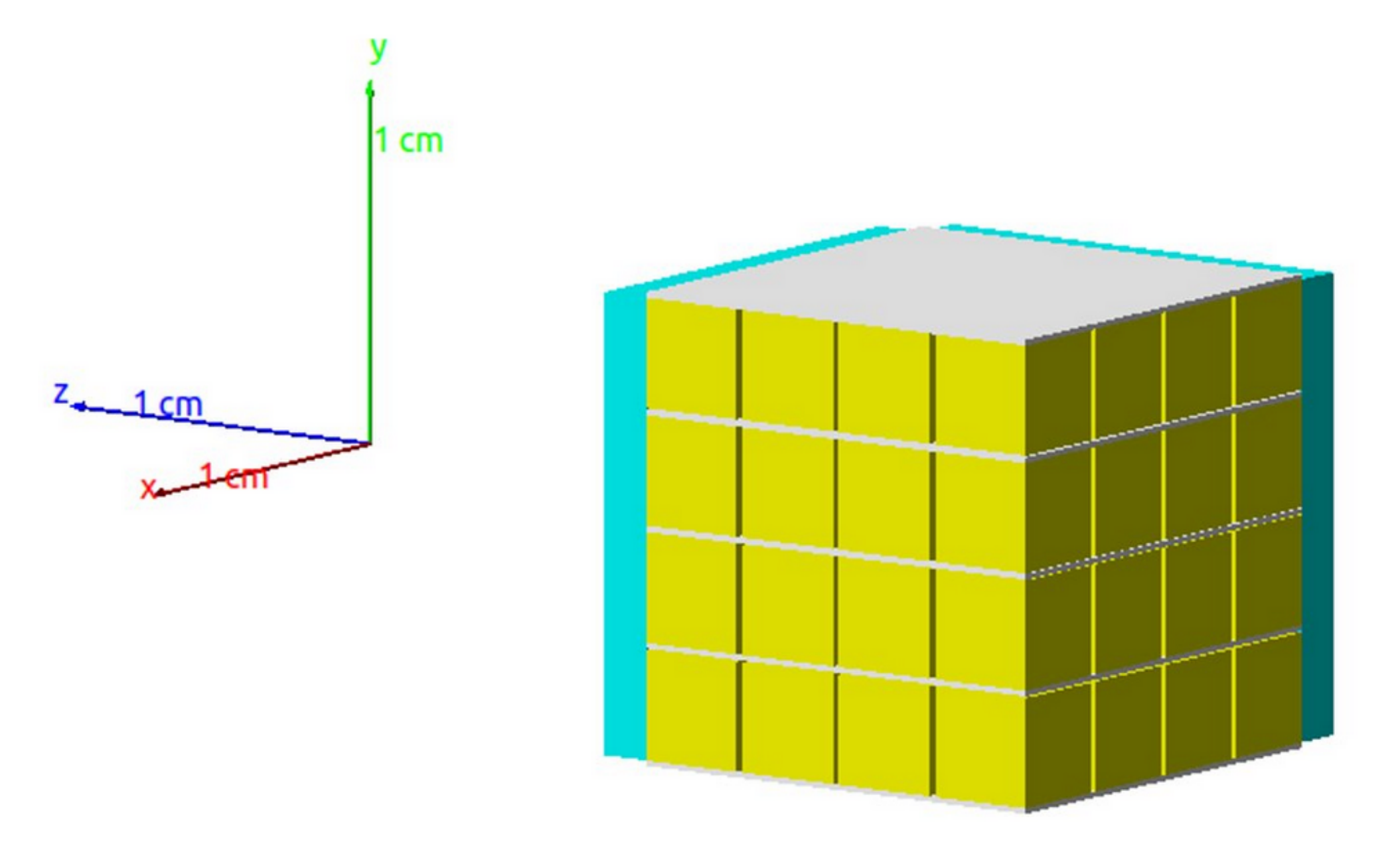}
  \caption{Visualisation of the simulated detector setup for the proposed omnidirectional radiation detector design. The voxels are GAGG crystals (in yellow), with reflectors (in white) and SiPMs (in cyan).}
  \label{fig:detector}
\end{figure}

We propose a novel type of gamma-ray detector with depth of interaction (DOI) capability, comprised of 64 scintillating crystals, as depicted in Fig. 1. The crystals are arranged in a 4x4x4 voxel array in a cubical formation, with crystal sizes of 3x3x3 mm$^3$ and a 3.2 mm matrix pitch. The read-out consists of two silicon photomultipliers (SiPMs), positioned perpendicularly to each other on adjacent faces of the segmented detector cube, with 1:1 coupling to the crystals. The SiPMs are 4x4 matrices of pixels, with pixel sizes of 3x3 mm$^2$ and a 3.2 mm pitch, which correspond to the sizes of the respective voxels to which they are attached. Reflectors, modelled as titanium dioxide (Ti0$_2$), are placed in planes that are perpendicular to both SiPMs, in between the 4x4 layers of the crystals. The reflectors provide maximal light collection by the SiPMs, as well as lower the possibility of inter-crystal light leakage from one plane to another. Additionally, reflectors encompass the outer faces of the detectors that are not in contact with the SiPMs. 

\subsection{Data acquisition and analysis}

Monte Carlo simulations of the proposed detector setups were developed in the Geant4 programming package, v11.3.2 \cite{AGOSTINELLI2003250}. To simulate the physics processes, the $\mathrm{QGSP\_BIC\_HP\_EMZ}$ physics list was used. This physics list includes the Doppler broadening phenomenon, including the momentum of electrons during a Compton scattering interaction. The GAGG crystals, i.e. the voxels, were selected as sensitive volumes in which the gamma rays deposit their energy. Point sources of relevant energies were placed in a spherical array with radius $r$ around the detectors, with coinciding centers of the detector and the sphere, so that:

\begin{equation}
  r = \frac{a\sqrt3}{2}\;+\;\mathrm{distance}.
  \label{eqn:chordlength}
\end{equation}

where $a$ is the length of the side of the detector ($a$ = 12.6 mm), and distance is a parameter of separation between the face or the vertex and the source. The output of the simulation included the calculation of the efficiencies for events in which the total deposited energy in the module exceeds 10 keV, as well as the information on the total energy deposited per event per voxel.

The total energy deposit in each voxel was smeared via Gaussian smearing on an event-by-event basis to achieve an energy spectrum expected in realistic conditions. The standard deviation of the Gaussian was chosen to be:
\begin{equation}
    \sigma(E) = \sqrt{\frac{E}{511}}\times \sigma_{511},
\label{eq:gauss}
\end{equation}
where $\sigma_{511}$ is deduced from the experimentally obtained energy resolution of 12\% full width at half-maximum (FWHM) in a similar setup \cite{Makek2020_crystals}, with energy values expressed in keV. The smearing resulted in energy resolutions of 10.5\% (FWHM) at 662 keV (corresponding to the experimentally observed values in \cite{DASH2023167808}) and 7.6\% (FWHM) at 1275 keV, preserving the $\sim 1/\sqrt{E}$ scaling of the energy resolution. 

Both data collection and analysis were optimized for utilization of the bash parallel command to expedite the computing processes \cite{ref-parallel}. The analysis of the collected data was conducted in ROOT \cite{root}. Plotting was done in Python, utilizing the Claude Opus 4.8 tool for drafting the data visualisation code.

\subsection{Detection efficiency calculations}

\begin{figure}
  \centering
  \begin{subfigure}[b]{0.45\textwidth}
    \centering
    \includegraphics[width=\textwidth]{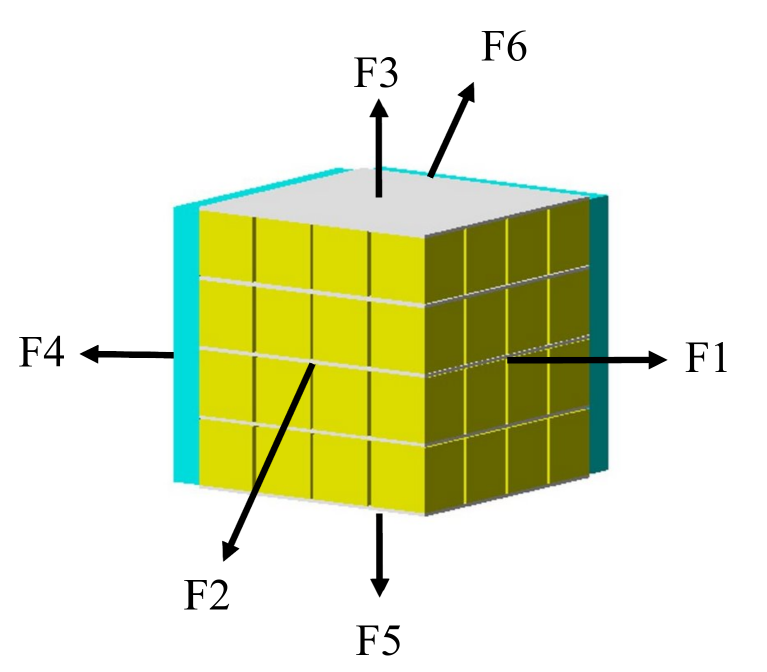}
    \caption{}
    \label{fig:faces}
  \end{subfigure}
  \begin{subfigure}[b]{0.45\textwidth}
    \centering
    \includegraphics[width=\textwidth]{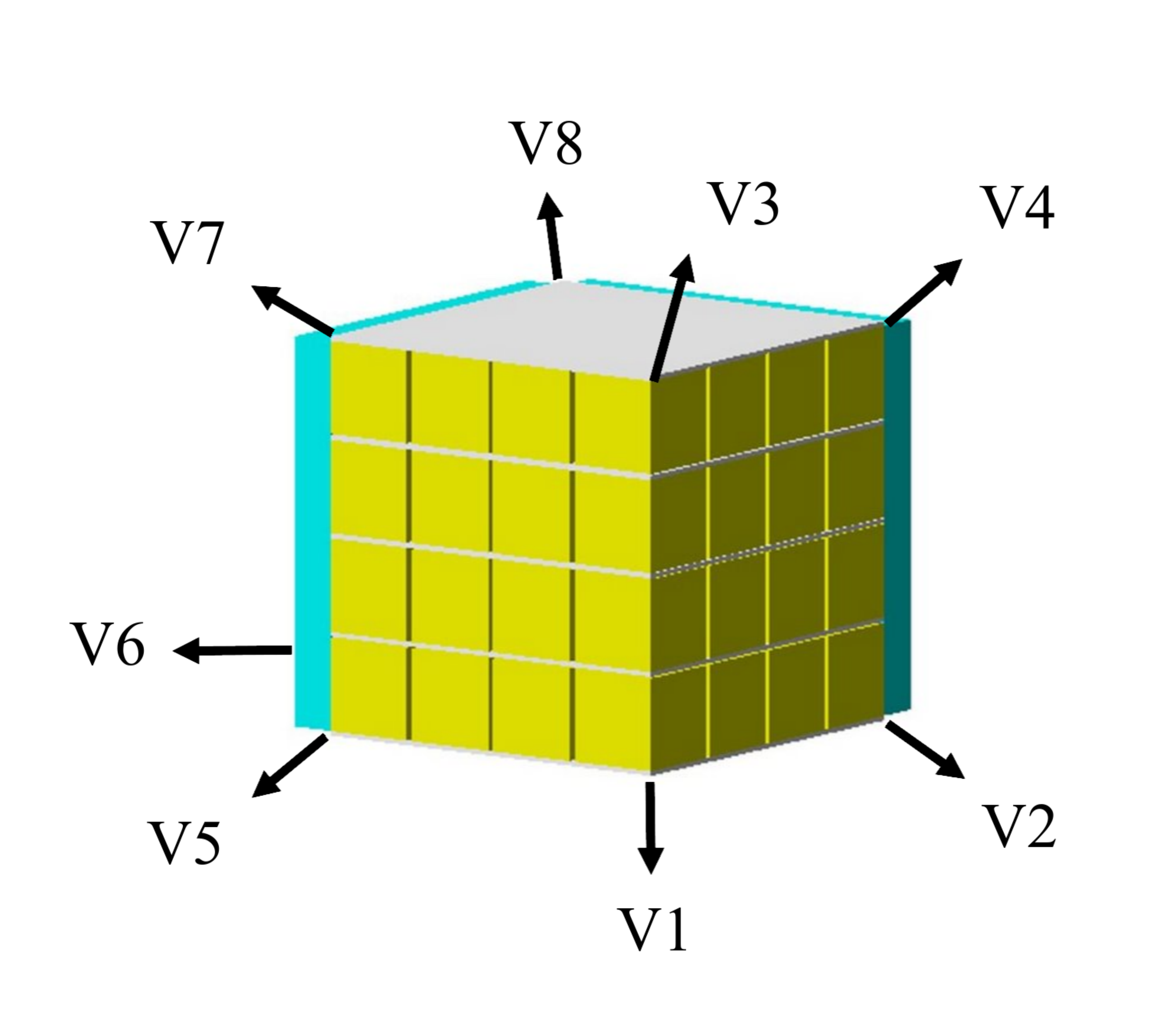}
    \caption{}
    \label{fig:vertices}
  \end{subfigure}
\caption{The faces (\protect\subref{fig:faces}) and vertices (\protect\subref{fig:vertices}) of the proposed detector.}
\label{fig:facesnvertices}
\end{figure}

The proposed detector design has a non-uniform geometry that could influence its response, as its sensitive volume is covered from two sides. To discern the detector’s response to the gamma rays from various point sources, intrinsic, geometric, and absolute efficiencies were calculated for each face and vertex of the detectors. 5 million events were utilized for these calculations.

Intrinsic efficiency $\epsilon\mathrm{_{int}}$ of a detector is defined as the ratio of the number of gamma rays that are detected ($N_{detected}$), and the number of gamma rays that enter the detector ($N_{enter}$) (Eq. \ref{eqn:intrinsic_mean}):

\begin{equation}
  \epsilon\mathrm{_{int}} = \frac{N_{detected}}{N_{enter}},
  \label{eqn:intrinsic_mean}
\end{equation}

with its error $\sigma\mathrm{_{int}}$ calculated as in Eq. \ref{eqn:intrinsic_sig}:

\begin{equation}
  \sigma\mathrm{_{int}} = \sqrt\frac{\epsilon\mathrm{_{int}}(\epsilon\mathrm{_{int}}-1)}{N_{enter}}.
  \label{eqn:intrinsic_sig}
\end{equation}
The intrinsic efficiency depends on the energy of the gamma rays that enter the detector, and the material and the thickness of the detector module.

The geometric efficiency $\epsilon\mathrm{_{geom}}$ of a detector is a measure of how well the detector intercepts radiation emitted from the source, and is defined as a number of gamma rays that enter the detector $N_{enter}$ divided by the number of gamma rays emitted from the source $N_{emitted}$ (Eq. \ref{eqn:geometric_mean}):
\begin{equation}
  \epsilon\mathrm{_{geom}} = \frac{N_{enter}}{N_{emitted}\;\mathrm{in\;4\pi}},
  \label{eqn:geometric_mean}
\end{equation}
with the error $\sigma\mathrm{_{geom}}$ (Eq. \ref{eqn:geometric_sig}) calculated assuming Poisson statistics:
\begin{equation}
  \sigma\mathrm{_{geom}} = \frac{\sqrt{N_{enter}}}{N_{emitted}\;\mathrm{in\;4\pi}}.
  \label{eqn:geometric_sig}
\end{equation}

The absolute efficiency of a detector $\epsilon\mathrm{_{abs}}$ is defined as the ratio of the number of photons that interacted with the detector $N_{detected}$ and the number of photons emitted by the source $N_{emitted}$ (Eq. \ref{eqn:absolute_mean}):
\begin{equation}
  \epsilon\mathrm{_{abs}} = \frac{N_{detected}}{N_{emitted}\;\mathrm{in\;4\pi}},
  \label{eqn:absolute_mean}
\end{equation}
and its error  $\sigma\mathrm{_{abs}}$ (Eq. \ref{eqn:absolute_sig}) 
\begin{equation}
  \sigma\mathrm{_{abs}} = \frac{\sqrt{N_{detected}}}{N_{emitted}\;\mathrm{in\;4\pi}}.
  \label{eqn:absolute_sig}
\end{equation}

The absolute efficiency measures the overall performance of the detection system. It is also related to the intrinsic $\epsilon\mathrm{_{int}}$ and geometric $\epsilon\mathrm{_{geom}}$ efficiencies, as it can be calculated from the relation in Eq. \ref{eqn:abs_total}:

\begin{equation}
  \epsilon\mathrm{_{abs}} = \epsilon\mathrm{_{int}}\times\epsilon\mathrm{_{geom}}.
  \label{eqn:abs_total}
\end{equation}

\subsection{Machine learning methods}

A machine learning (ML) model was developed with the XGBoost library \cite{xgboost}, which utilizes supervised learning algorithms based on gradient-boosted decision trees. It is available as an application programming interface (API) in Python's scikit-learn library \cite{scikit-learn}. XGBoost library is widely used for ML tasks such as classification and regression, since it is designed to be highly efficient, portable, and flexible. The draft of the code for utilization of the XGBoost library was developed using Claude Opus 4.8.

\begin{figure}[t]
  \centering
  \includegraphics[width=\linewidth]{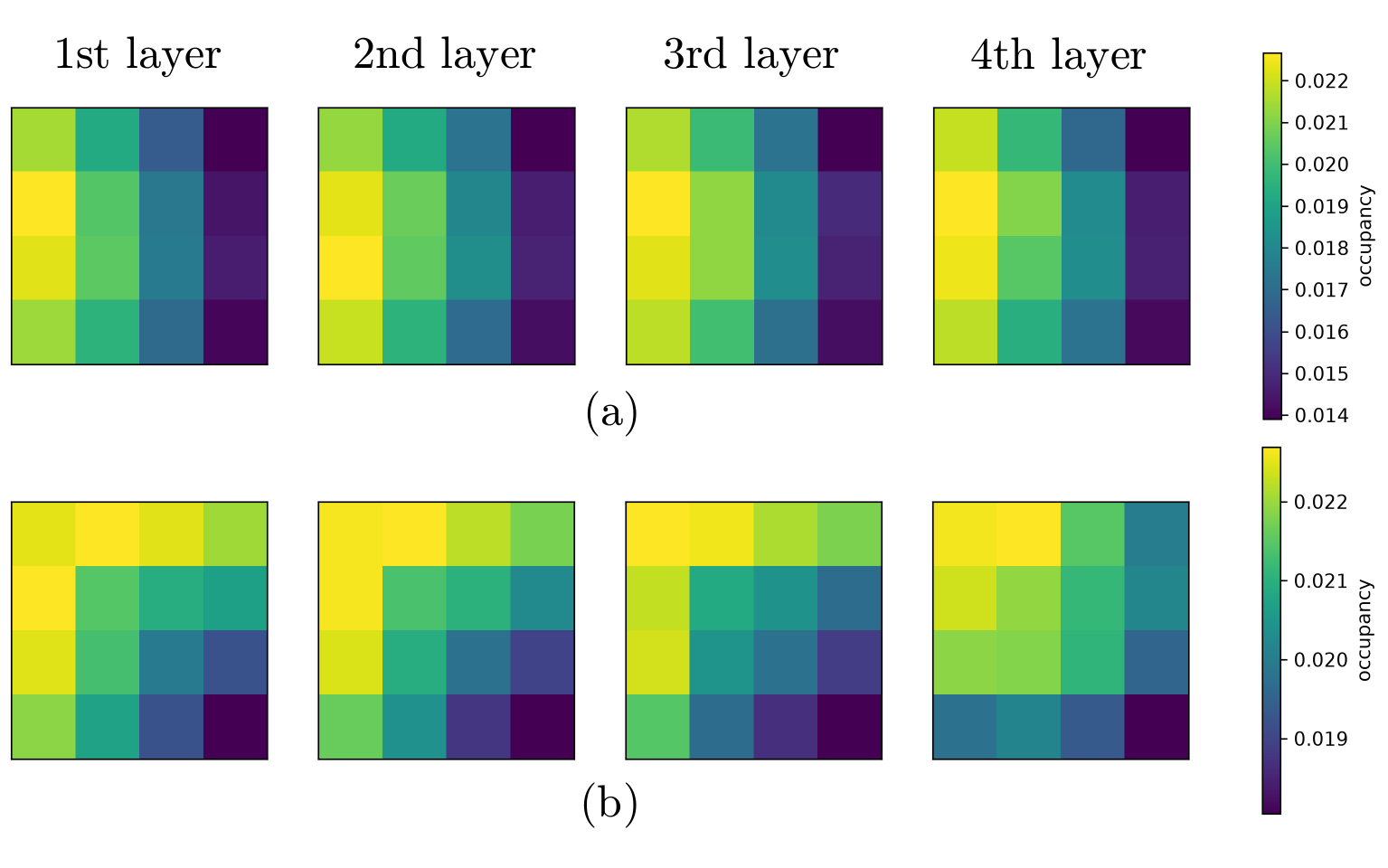}
  \caption{The occupancy of the voxels in a 2$\times$10$^6$ sample of events obtained for a 662 keV source positioned 100 mm in front of the face F2 in (a), and in front of the V5 vertex (b). }
  \label{fig:occupancy}
\end{figure}

As the point sources of the 511, 662, and 1275 keV energies were positioned around the detector at various distances, there were 924 classes from:
\begin{itemize}
    \item 3 energies (511, 662 and 1275 keV),
    \item 14 positions in front of faces and vertices,
    \item and 22 distances at each position, ranging from 10 mm to 300 mm.
\end{itemize}
The grid spacing of the training distances from 10 mm to 100 mm was 10 mm, from 100 mm to 200 mm it was 12.5 mm, and from 200 mm to 300 mm it was 25 mm. 
The features chosen for training were:
\begin{itemize}
    \item occupancy of each of the 64 detector voxels, i.e., the fractions of events that fired in each voxel,
    \item photopeak fraction at $\pm2\sigma$ from the peak,
    \item 3 multiplicity fractions: 1-voxel, 2-voxel and (3+)-voxel events,
    \item log-rate of the number of collected events,
\end{itemize}
resulting in 132 features in total. The occupancy of the voxels will vary during irradiation from different source positions, as the gamma rays attenuate in the detector, and its effect can be seen in Figure \ref{fig:occupancy}. The energy depositions are known, as it is assumed that the detector is calibrated. The multiplicity fractions slightly differ between faces and vertices, and the number of collected events, represented by the log-rate, will depend on the energy and position of the source, but also on its distance from the detector.

The total of 50 million events per class was separated into three categories: 60\% for training, 20\% for validation, and 20\% for testing. After Gaussian smearing of the energy resolution (Eq. \ref{eq:gauss}), an energy cutoff at 100 keV was applied to correspond to the realistic experimental conditions of background noise removal. The pseudomeasurement grouping was defined as 200 000 lines per class. The XGBoost ML model's regression capabilities were utilized for the deduction of the source's distance from the face or the vertex in front of which it was positioned. The hyperparameters relevant for the model's training are summarized in Table \ref{tab:params}. The regression was performed on the logarithm of the Distance (log$_{10}$(Distance)) to minimize the errors at large Distances, and transformed back after regression. 

\begin{table}
\centering
\small
\caption{XGBoost hyperparameters used in model training.}
\label{tab:params}
\begin{tabular}{|c|c|c|}
\hline
    Hyperparameter & Meaning & Regression \\ \hline
    n\_estimators     & Max number of trees in the ensemble  & 5000 \\ 
    max\_depth        & Max depth of the trees & 3 \\
    learning\_rate    & Step size of each boosting iteration & 0.07   \\ 
    subsample         & Fraction of samples used for training & 0.8    \\ 
    colsample\_bytree & Fraction of features used for training & 0.5   \\
    tree\_method      & Tree construction algorithm & hist  \\ 
    objective         & Learning task and loss function & reg:squarederror   \\
    eval\_metric      & Evaluation of model performance metric & MAE   \\
    early\_stopping\_rounds & Number of rounds with no improvement &  30 \\
    random\_state     & Random number seed & 42   \\
    \hline
\end{tabular}
\end{table}

\subsection{ML model's performance evaluation and validation}
To assess the accuracy of the model's prediction capability, three measures were used: Mean Absolute Error (MAE), Root Mean Square Error (RMSE), and the coefficient of determination R$^2$. The MAE represents the average magnitude of the errors made by a predictive model and is calculated as:
\begin{equation}
  \mathrm{{MAE}} = \frac{\sum_{i=1}^{n}|y_{observed, i}-y_{predicted,i}|}{n}.
  \label{eqn:mae}
\end{equation}
The RMSE value is calculated by:
\begin{equation}
  \mathrm{{RMSE}} =\sqrt \frac{\sum_{i=1}^{n}|y_{observed, i}-y_{predicted,i}|^2}{n},
  \label{eqn:rmse}
\end{equation}
and evaluates the prediction potential of the model, but unlike MAE, it weights larger errors more heavily than the smaller ones. The R$^2$ metric is defined as:
\begin{equation}
  \mathrm{{R^2}} =1 - \frac{\sum_{i=1}^{n}|y_{observed, i}-y_{predicted,i}|^2}{\sum_{i=1}^{n}|y_{observed, i}-\bar{y}_{observed,i}|^2},
  \label{eqn:r2}
\end{equation}
and describes how much variance in the target variable is explained by the model.

To validate predictive performance, the model was trained by excluding the target distances at each energy and then evaluated by interpolating the held-out data. The bias in the distance interpolation is calculated as the average of the difference between predicted and true values, testing the systematic offset in the predicted values. 

\section{Results}\label{sec:results}
\subsection{Detector's response due to the geometrical non-uniformity}

\begin{figure}
  \centering
  \begin{subfigure}[b]{\textwidth}
    \centering
    \includegraphics[width=\textwidth]{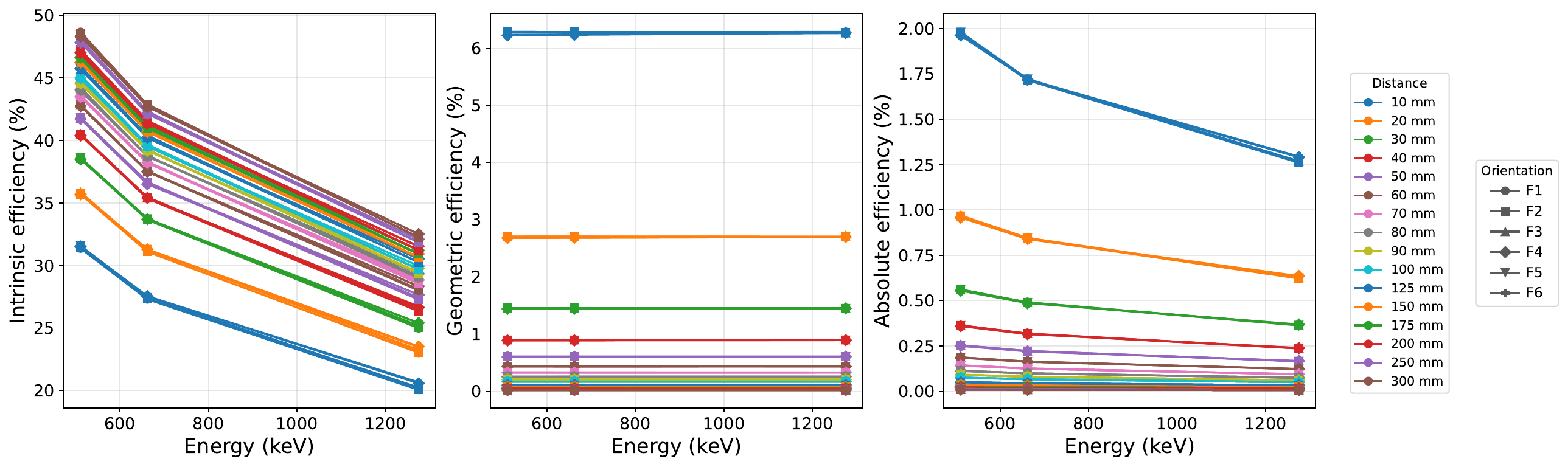}
    \caption{}
    \label{fig:energy_plots_1}
  \end{subfigure}
  \begin{subfigure}[b]{\textwidth}
    \centering
    \includegraphics[width=\textwidth]{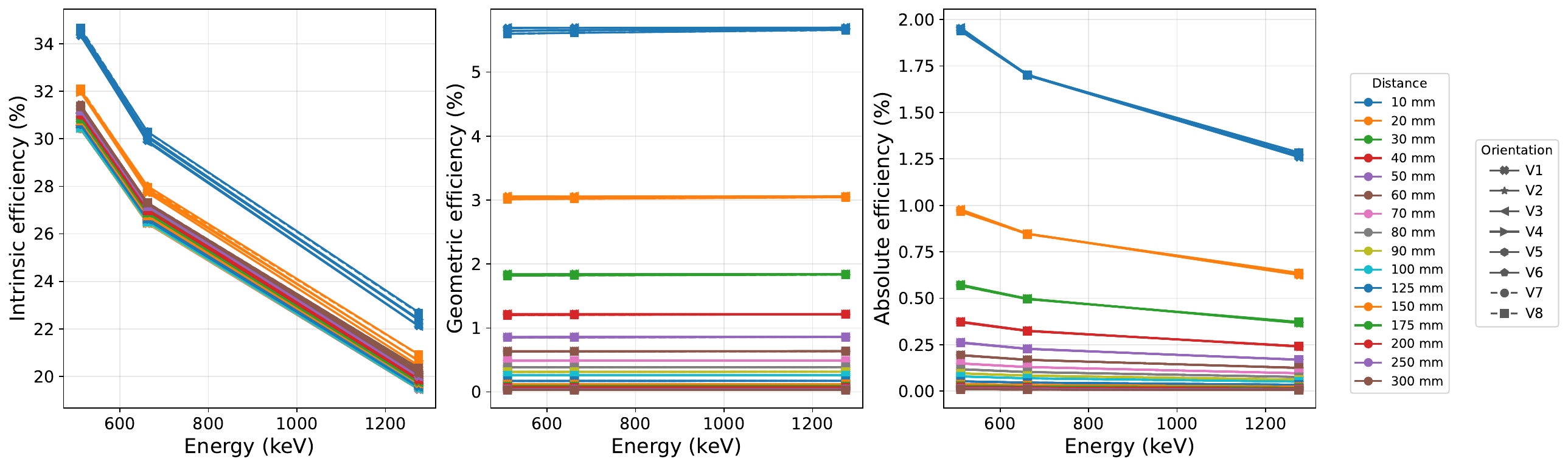}
    \caption{}
    \label{fig:energy_plots_2}
  \end{subfigure}
\caption{The $\epsilon\mathrm{_{int}}$, $\epsilon\mathrm{_{geom}}$ and $\epsilon\mathrm{_{abs}}$ dependence on the 511, 662 and 1275 keV energies at various distances for faces (\protect\subref{fig:energy_plots_1}) and vertices (\protect\subref{fig:energy_plots_2}) of the proposed detector.}
\label{fig:energy_plots}
\end{figure}

\begin{figure}
  \centering
  \begin{subfigure}[b]{\textwidth}
    \centering
    \includegraphics[width=\textwidth]{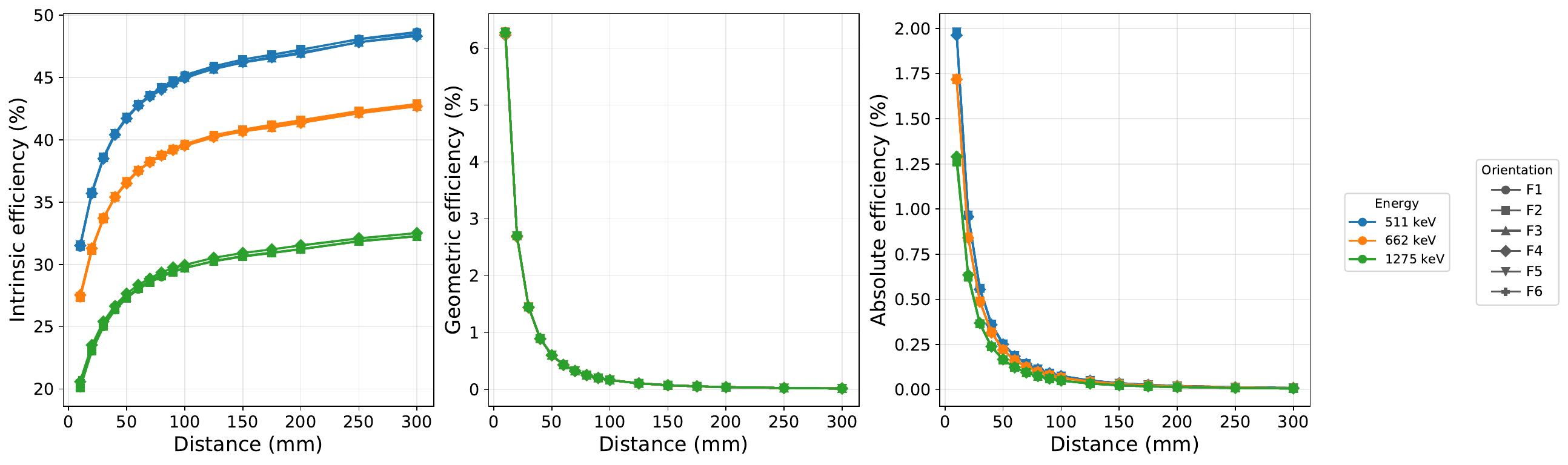}
    \caption{}
    \label{fig:distance_plots_1}
  \end{subfigure}
  \begin{subfigure}[b]{\textwidth}
    \centering
    \includegraphics[width=\textwidth]{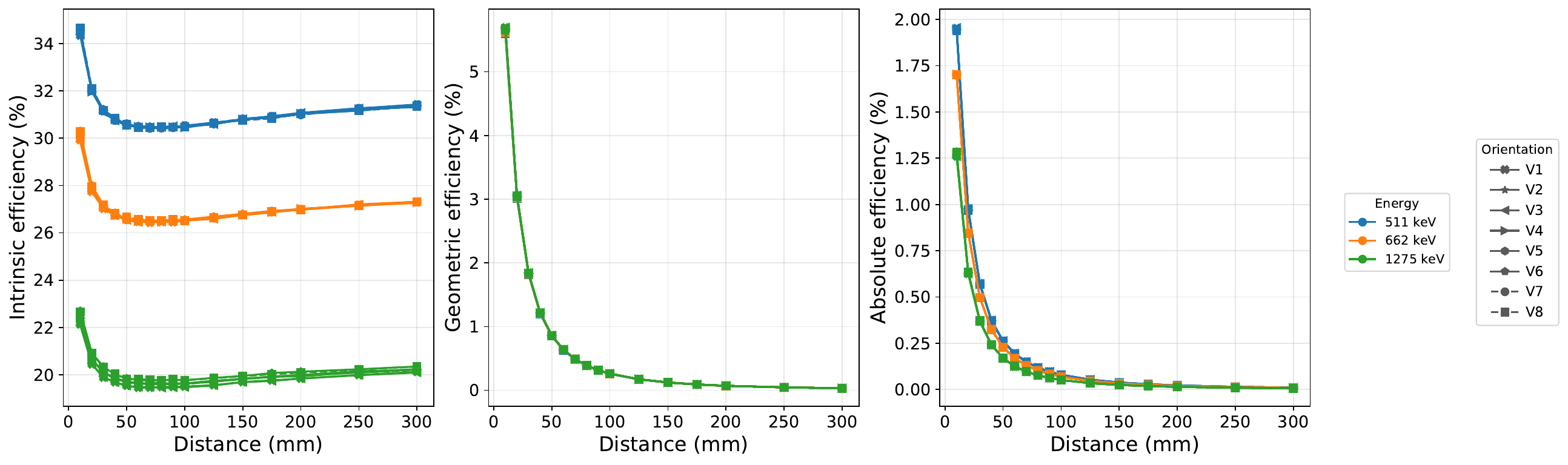}
    \caption{}
    \label{fig:distance_plots_2}
  \end{subfigure}
  \caption{The $\epsilon\mathrm{_{int}}$, $\epsilon\mathrm{_{geom}}$ and $\epsilon\mathrm{_{abs}}$ dependence on distance for 511, 662 and 1275 keV energies for faces (\protect\subref{fig:distance_plots_1}) and vertices (\protect\subref{fig:distance_plots_2}) of the proposed detector.}
  \label{fig:distance_plots}
\end{figure}

The proposed design of the detector relies on the SiPM read-out from two neighboring faces of the detector cube. The effect of this geometrical non-uniformity of the proposed detector system was studied by efficiency calculations of the gamma-ray photons that deposited more than 10 keV of their energy in one of the voxels in the detector module. The $\epsilon\mathrm{_{int}}$, $\epsilon\mathrm{_{geom}}$, and $\epsilon\mathrm{_{abs}}$ were measured and calculated for all of the detector’s faces and vertices at source energies corresponding to Na-22 and Cs-137 at 511, 662, and 1275 keV, at source-to-detector distances ranging from 10 mm to 300 mm. The results can be seen in Figs. \ref{fig:energy_plots} and \ref{fig:distance_plots}, comparing the efficiencies for the faces and vertices of the detector at the selected energies and distances, respectively. It can be observed that intrinsic efficiency lowers with increasing energy at every distance (Figs. \ref{fig:energy_plots_1} and \ref{fig:energy_plots_2}), for both faces and vertices. The intrinsic efficiency loss from 511 keV to 1275 keV point source energies is from $\sim$38\% for 10 mm source-to-detector distance to $\sim$33\% for 300 mm separation if the source is in front of the faces, while for the vertices it is approximately 37\% at every distance. The intrinsic efficiencies for 511, 662 and 1275 keV energies for faces (Fig. \ref{fig:distance_plots_1}) and vertices (Fig. \ref{fig:distance_plots_2}) at different source-to-detector distances show opposite tendencies. The faces exhibit an increase in intrinsic efficiencies as the distance increase from 10 mm to 300 mm: $\sim$50\% for 511 keV, $\sim$60\% for 662 keV, and $\sim$65\% for 1275 keV. Intrinsic efficiency values for the vertices in the same distance interval are lower by $\sim$11\% for 511 keV, $\sim$12\% for 662 keV, and $\sim$13\% for 1275 keV point source energies. Geometric efficiencies do not depend on the source energies, but change exponentially with the source-to-detector distance. The absolute efficiencies for point sources of different energies also exponentially decrease with distance, but for distances below 100 mm, the energies show slight differentiation, as the lowest energy of 511 keV has the highest absolute efficiency of up to 2\% for the closest positions, while the 662 keV and 1275 keV energies have lower efficiencies of 1.7\% and 1.25\%, respectively.

\subsection{Regression model of the source-to-detector distances}

\begin{figure}
  \centering
  \begin{subfigure}[b]{0.6\textwidth}
    \centering
    \includegraphics[width=\textwidth]{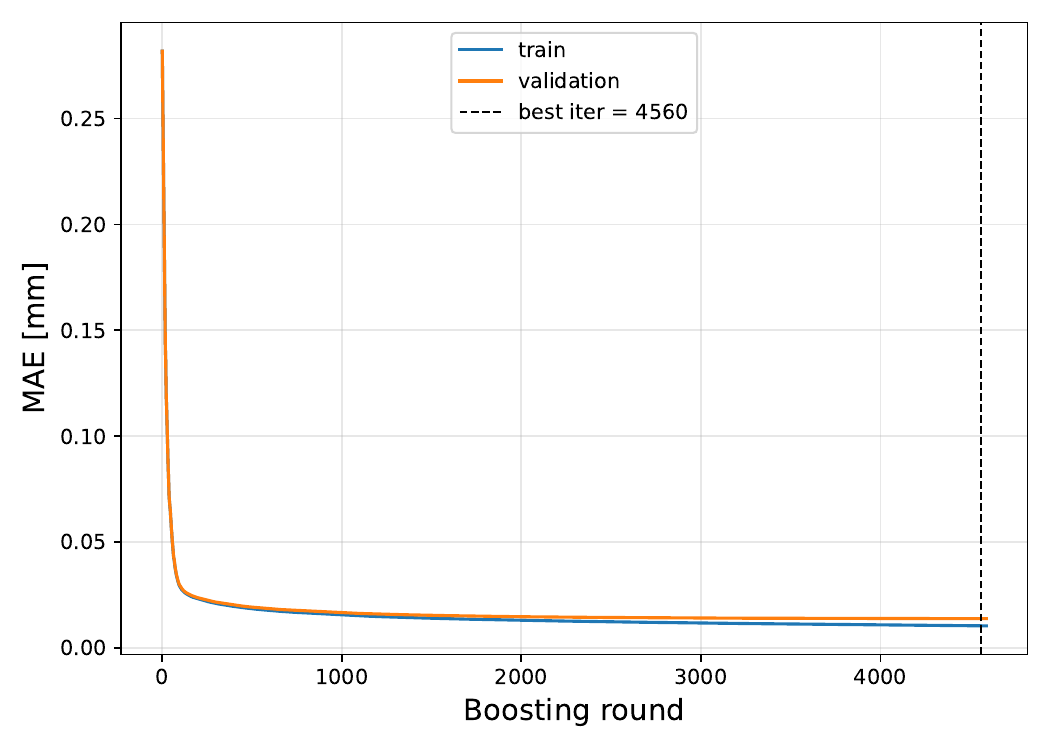}
    \caption{}
    \label{fig:curve_1}
  \end{subfigure}
  \begin{subfigure}[b]{0.6\textwidth}
    \centering
    \includegraphics[width=\textwidth]{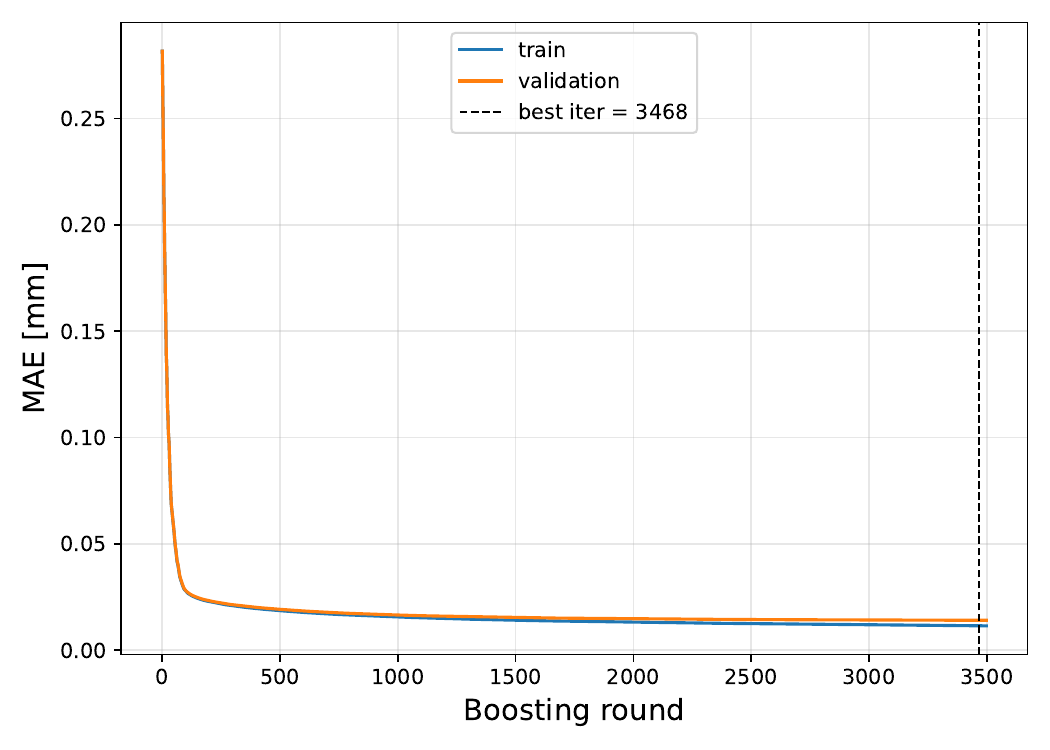}
    \caption{}
    \label{fig:curve_2}
    \end{subfigure}
  \begin{subfigure}[b]{0.6\textwidth}
    \centering
    \includegraphics[width=\textwidth]{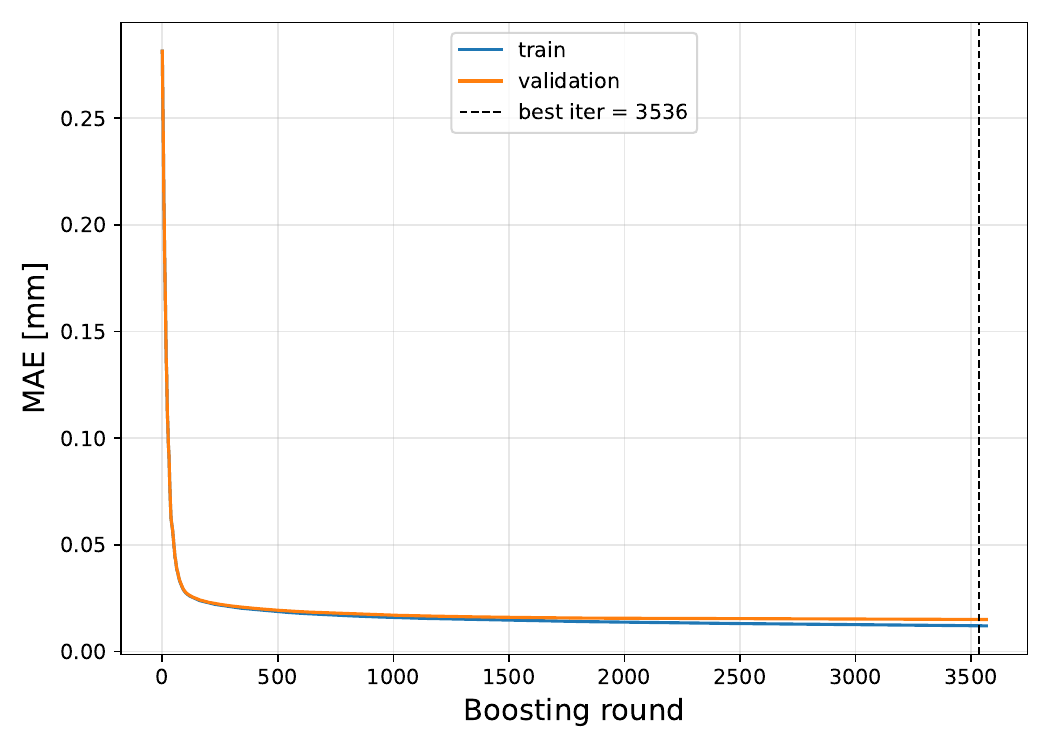}
    \caption{}
    \label{fig:curve_3}
  \end{subfigure}
  \caption{Comparison of the training and validation learning curves for XGBoost regression models at (\protect\subref{fig:curve_1}) 511, (\protect\subref{fig:curve_2}) 662, and (\protect\subref{fig:curve_3}) 1275 keV energies.}
  \label{fig:curve_plots}
\end{figure}

\begin{figure}
  \centering
  \begin{subfigure}[b]{0.45\textwidth}
    \centering
    \includegraphics[width=\textwidth]{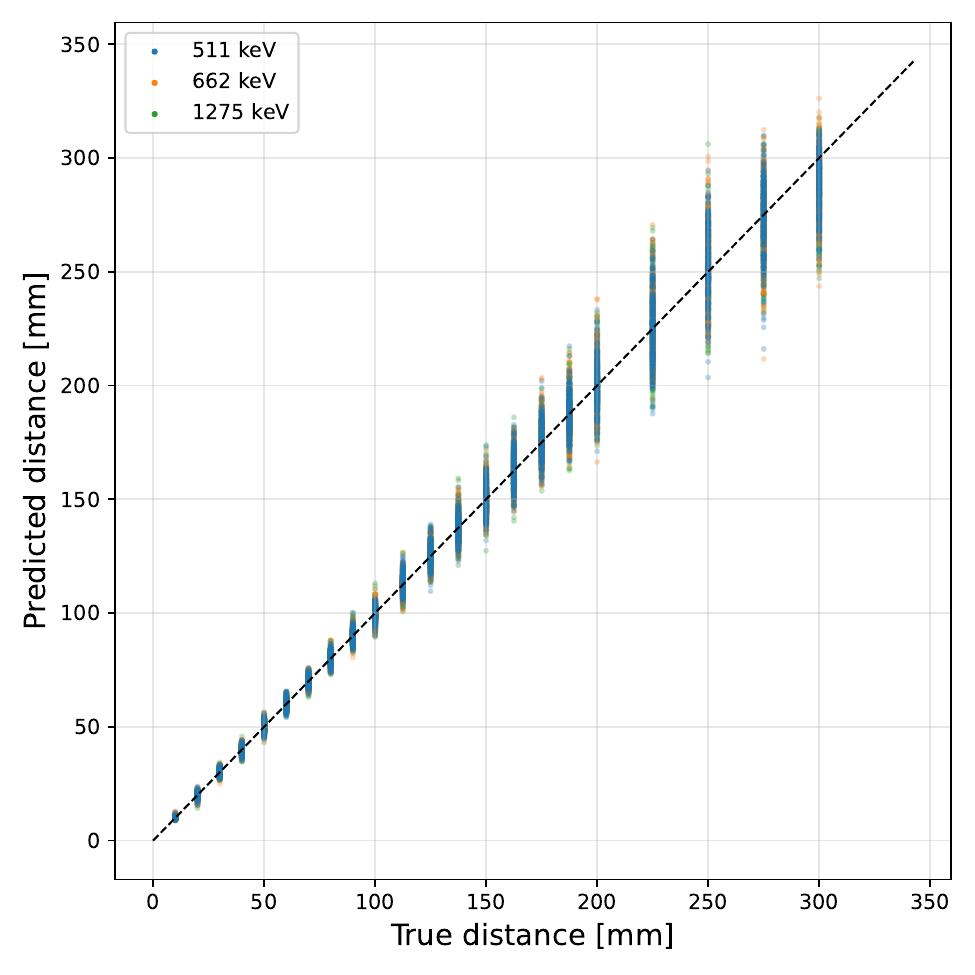}
    \caption{}
    \label{fig:reg_1}
  \end{subfigure}
  \begin{subfigure}[b]{0.45\textwidth}
    \centering
    \includegraphics[width=\textwidth]{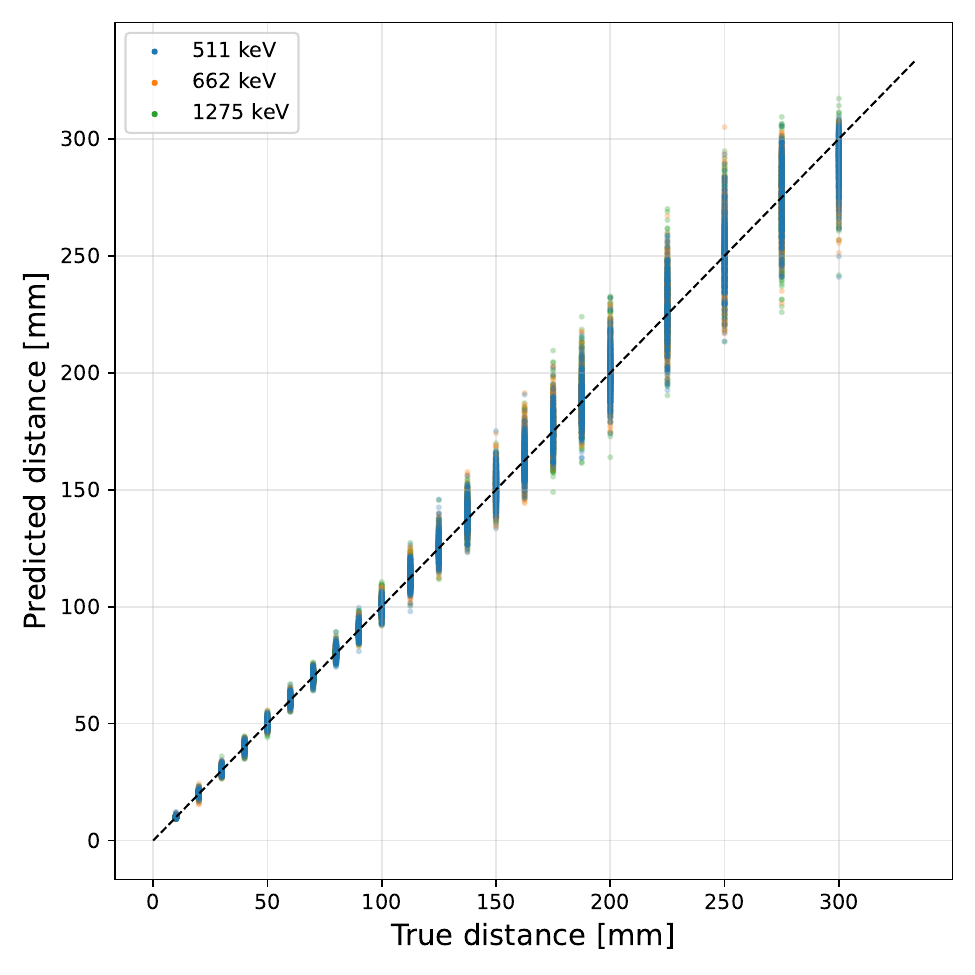}
    \caption{}
    \label{fig:reg_2}
  \end{subfigure}
  \caption{The distance regression results at 511, 662 and 1275 keV energies for (\protect\subref{fig:reg_1}) faces and (\protect\subref{fig:reg_2}) vertices of the proposed detector.}
  \label{fig:prediction_plots}
\end{figure}

\begin{figure}
  \centering
  \begin{subfigure}[b]{0.45\textwidth}
    \centering
    \includegraphics[width=\textwidth]{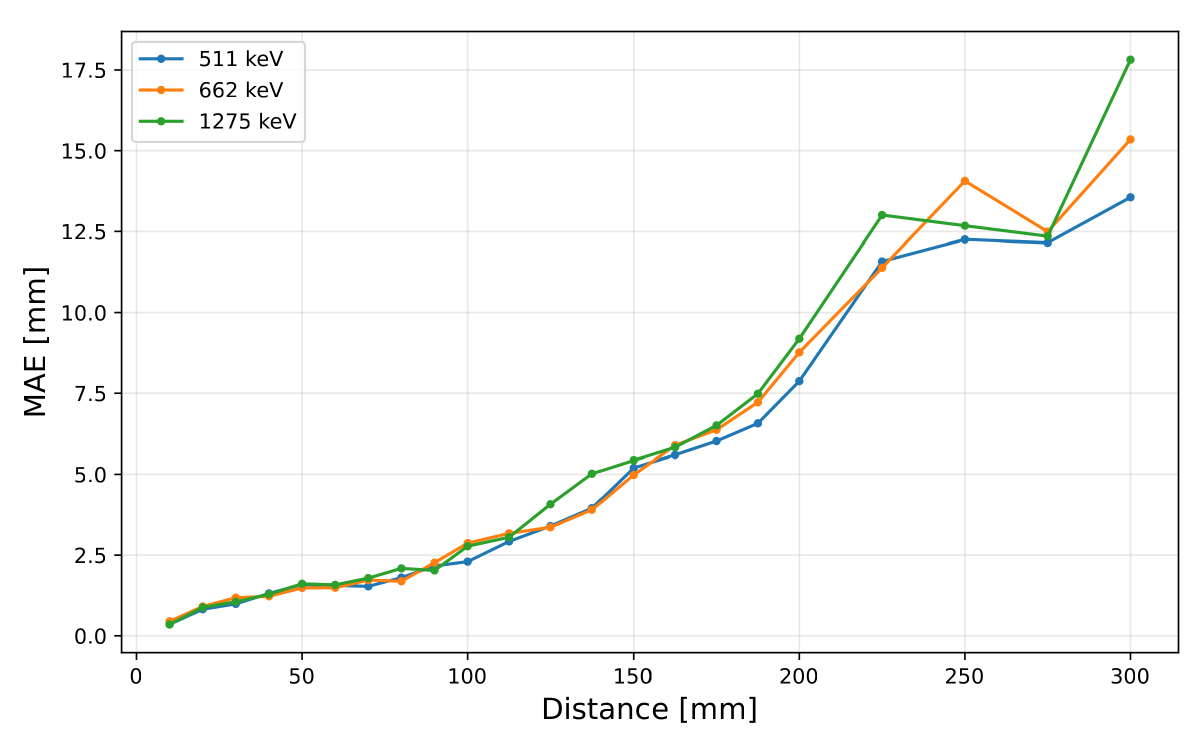}
    \caption{}
    \label{fig:maemodel_faces}
  \end{subfigure}
  \begin{subfigure}[b]{0.45\textwidth}
    \centering
    \includegraphics[width=\textwidth]{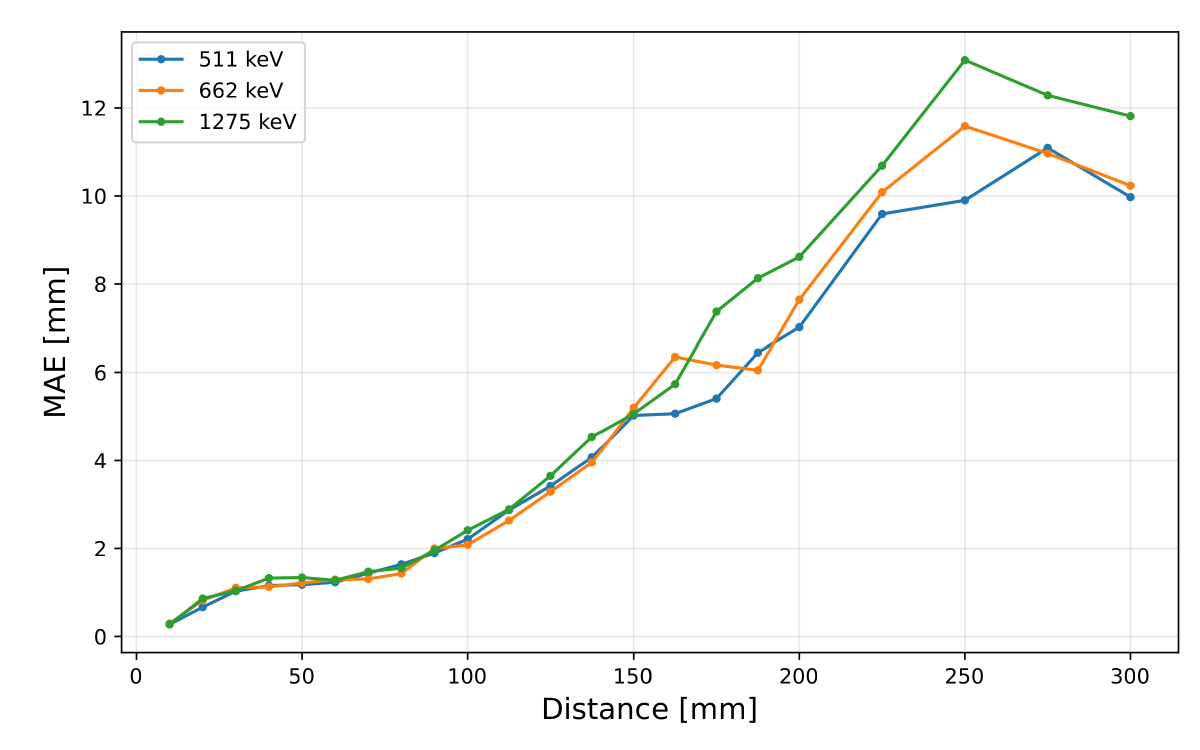}
    \caption{}
    \label{fig:maemodel_verts}
  \end{subfigure}
    \begin{subfigure}[b]{0.45\textwidth}
    \centering
    \includegraphics[width=\textwidth]{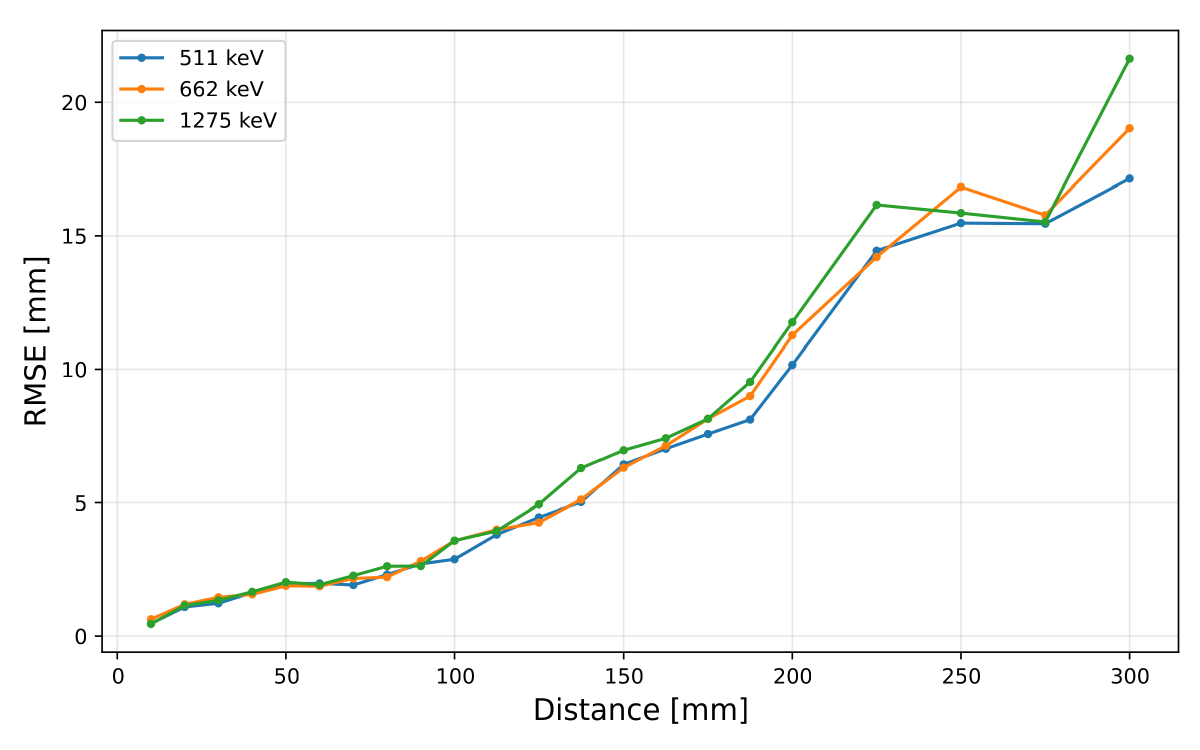}
    \caption{}
    \label{fig:rmsemodel_faces}
  \end{subfigure}
  \begin{subfigure}[b]{0.45\textwidth}
    \centering
    \includegraphics[width=\textwidth]{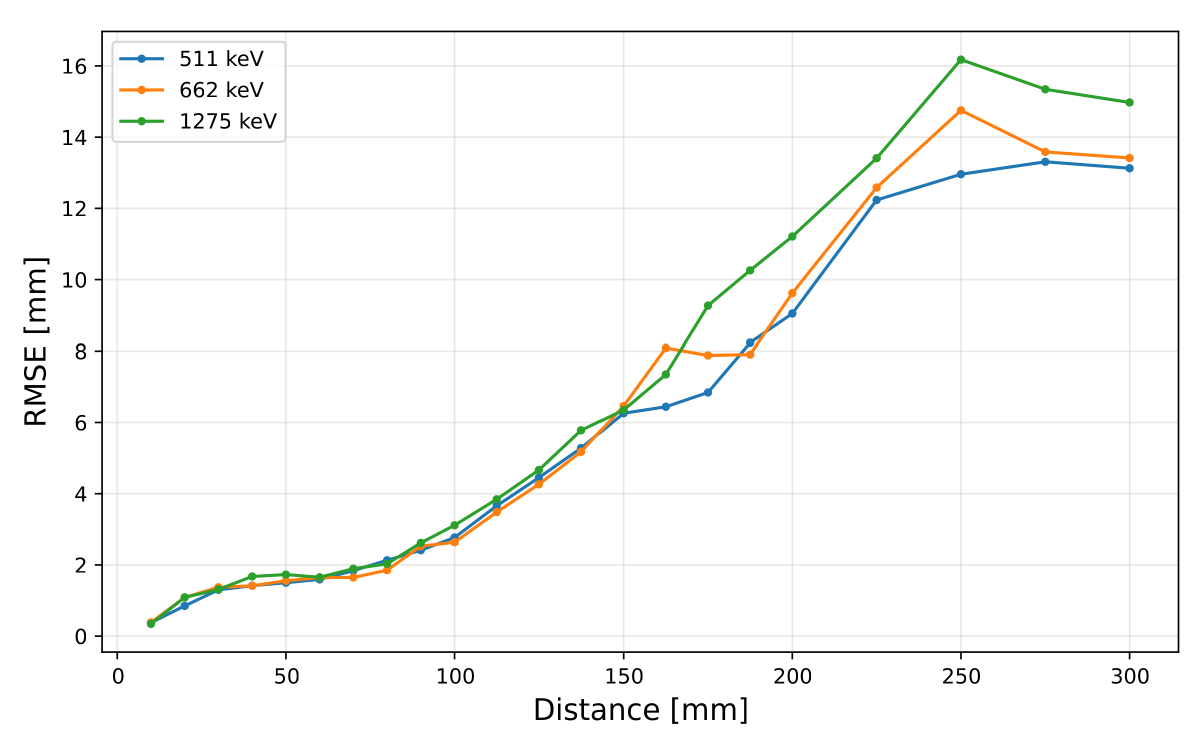}
    \caption{}
    \label{fig:rmsemodel_verts}
  \end{subfigure}
  \caption{The MAE and the RMSE values for models at 511, 662, and 1275 keV energies for faces (Figures (\protect\subref{fig:maemodel_faces}) and (\protect\subref{fig:rmsemodel_faces})) and vertices (Figures (\protect\subref{fig:maemodel_verts}) and (\protect\subref{fig:rmsemodel_verts})) of the proposed detector.}
  \label{fig:errorreg_plots}
\end{figure}

The regression models for source-to-detector distance determination were trained on all the available data, i.e., 50 million events per class for all 22 distances, for each point source energy (511, 662, and 1275 keV). The learning curves obtained during training and validation for the models are compared in Fig. \ref{fig:curve_plots}. The curves are well matched, demonstrating the absence of overfitting in the models and their stable convergence. The results of the regression for the predicted distances vs. their true values are presented in Figure \ref{fig:prediction_plots}. The most useful feature for the models' training was log-rate, with importance scores of 0.099 for the 511 keV model, 0.111 for the 662 keV model, and 0.106 for the 1275 keV model. It is evident that the model's accuracy is highest for low source-to-detector separation, and gradually falls off with increasing distance. The values in Table \ref{tab:regression} represent the average MAEs and RMSEs in the model's predictions across all distances. The results show that the errors are increasing with the point source energy, but their mean remains below the lowest grid spacing, suggesting that the models have strong prediction capabilities. In Fig. \ref{fig:errorreg_plots}, MAEs and RMSEs are presented for every distance on which the models were trained. As the regression results in Fig. \ref{fig:prediction_plots} indicate, the errors are higher for larger distances and grid spacings. The values of MAEs range from less than 2.5 mm for distances up to 100 mm (with 10 mm grid spacing) for faces and vertices, to $\sim$13 mm for faces and vertices at larger distances (Fig. \ref{fig:maemodel_faces} and \ref{fig:maemodel_verts}). The RMSEs follow the same trend as MAEs, with a slightly steeper rise from the low to high distances, as their calculation penalizes larger errors more heavily, with $\sim$15 mm RMSE for 250 mm distances (25 mm grid spacing) from the faces (Fig. \ref{fig:rmsemodel_faces}), and $\sim$14 mm RMSE for the vertices (Fig. \ref{fig:rmsemodel_verts}). The R$^2$ values in Table \ref{tab:regression} show that the models fit the data almost perfectly, with approximately 99\% of the variation in the data explained by the models.

\begin{table}
\centering
\caption{XGBoost regression model results for every energy on all distances from faces and vertices of the detector.}
\label{tab:regression}
\begin{tabular}{|c|c|c|c|}
\hline

    \multicolumn{4}{|c|}{} \\
    \multicolumn{4}{|c|}{TOTAL} \\
    \multicolumn{4}{|c|}{} \\\hline
    Energy [keV] & MAE [mm] & RMSE [mm] & R$^2$ [\%]\\ \hline
    511          & 4.5      &    7.4    &   0.9921 \\
    662          & 4.7      &    7.8    &   0.9911   \\
    1275         & 5.1      &    8.4    &   0.9896   \\
    \hline\hline
    \multicolumn{4}{|c|}{} \\
    \multicolumn{4}{|c|}{FACES} \\
    \multicolumn{4}{|c|}{} \\\hline
    Energy [keV] & MAE [mm] & RMSE [mm] & R$^2$ \\ \hline
    511          & 4.8      &    8.0    &   0.9921 \\
    662          & 5.1      &    8.4    &   0.9896   \\
    1275         & 5.4      &    8.9    &   0.9884   \\
    \hline\hline
    \multicolumn{4}{|c|}{} \\
    \multicolumn{4}{|c|}{VERTICES} \\
    \multicolumn{4}{|c|}{} \\\hline
    Energy [keV] & MAE [mm] & RMSE [mm] & R$^2$ \\ \hline
    511          & 4.2      &    6.9    &   0.9931 \\
    662          & 4.4      &    7.2    &   0.9923   \\
    1275         & 4.9      &    8.0    &   0.9905   \\
    \hline
\end{tabular}
\end{table}

\subsection{Validation of the regression model}\label{sec:validation}

\begin{figure}
  \centering
  \begin{subfigure}[b]{0.45\textwidth}
    \centering
    \includegraphics[width=\textwidth]{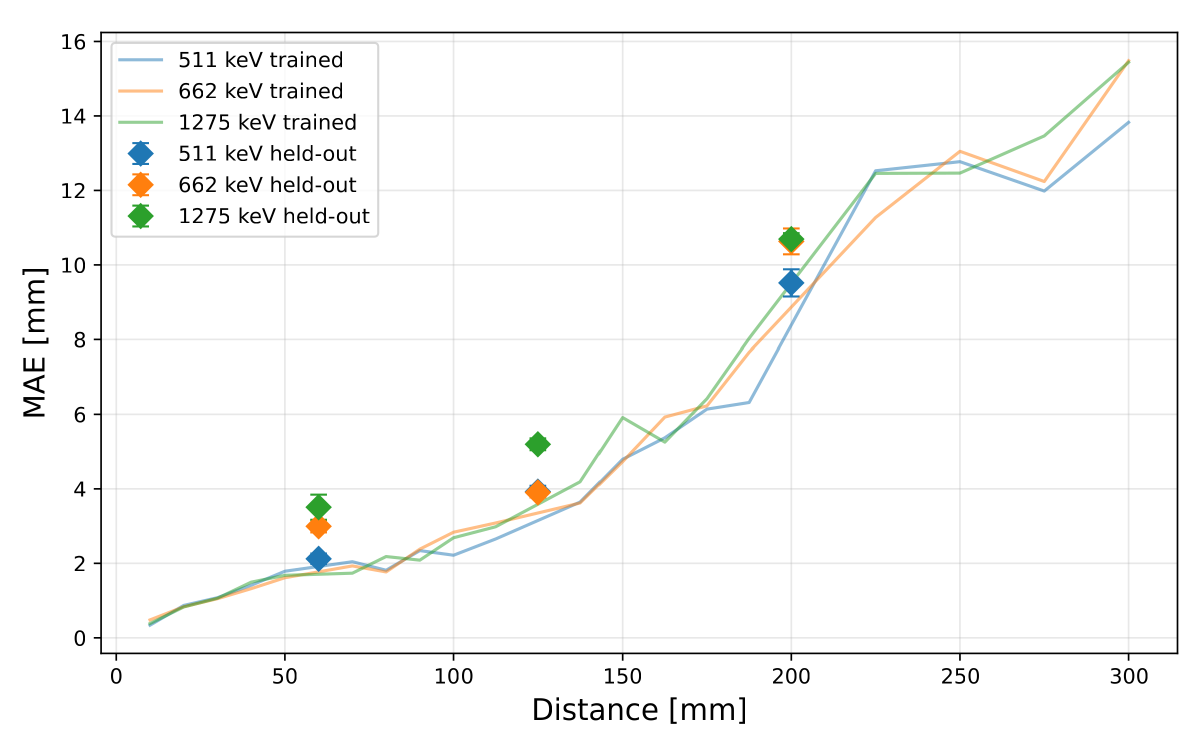}
    \caption{}
    \label{fig:mae_faces}
  \end{subfigure}
  \begin{subfigure}[b]{0.45\textwidth}
    \centering
    \includegraphics[width=\textwidth]{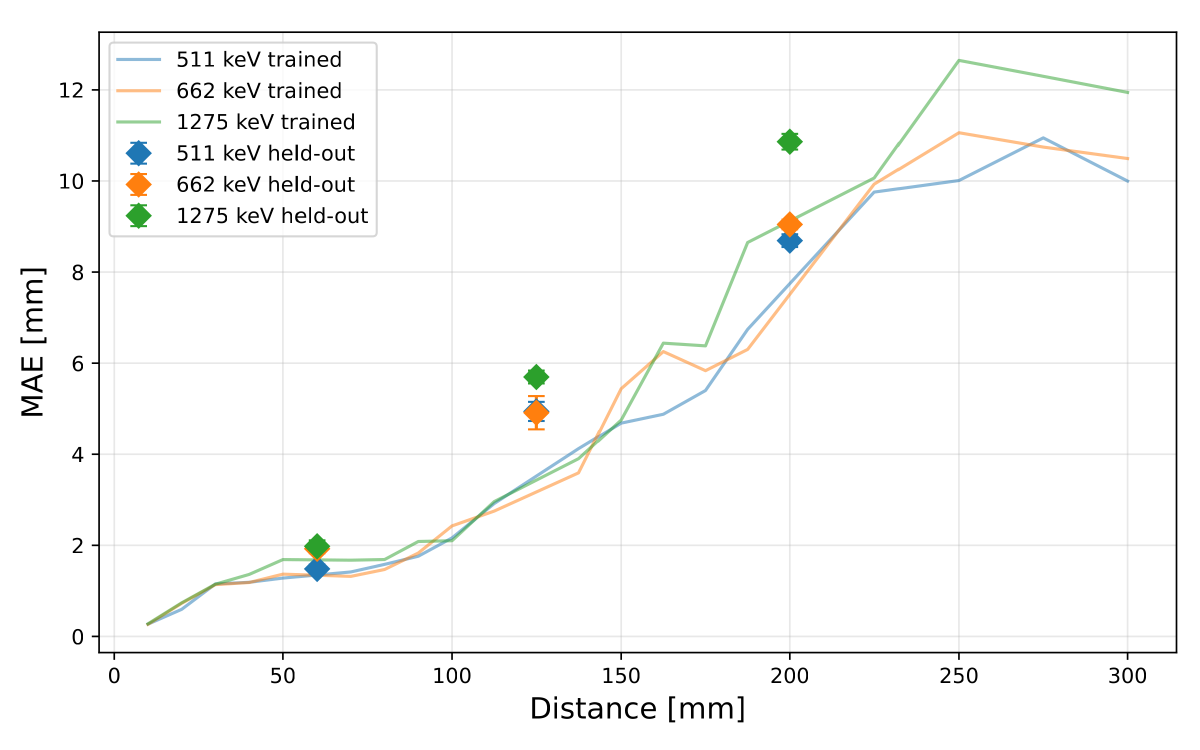}
    \caption{}
    \label{fig:mae_verts}
  \end{subfigure}
    \begin{subfigure}[b]{0.45\textwidth}
    \centering
    \includegraphics[width=\textwidth]{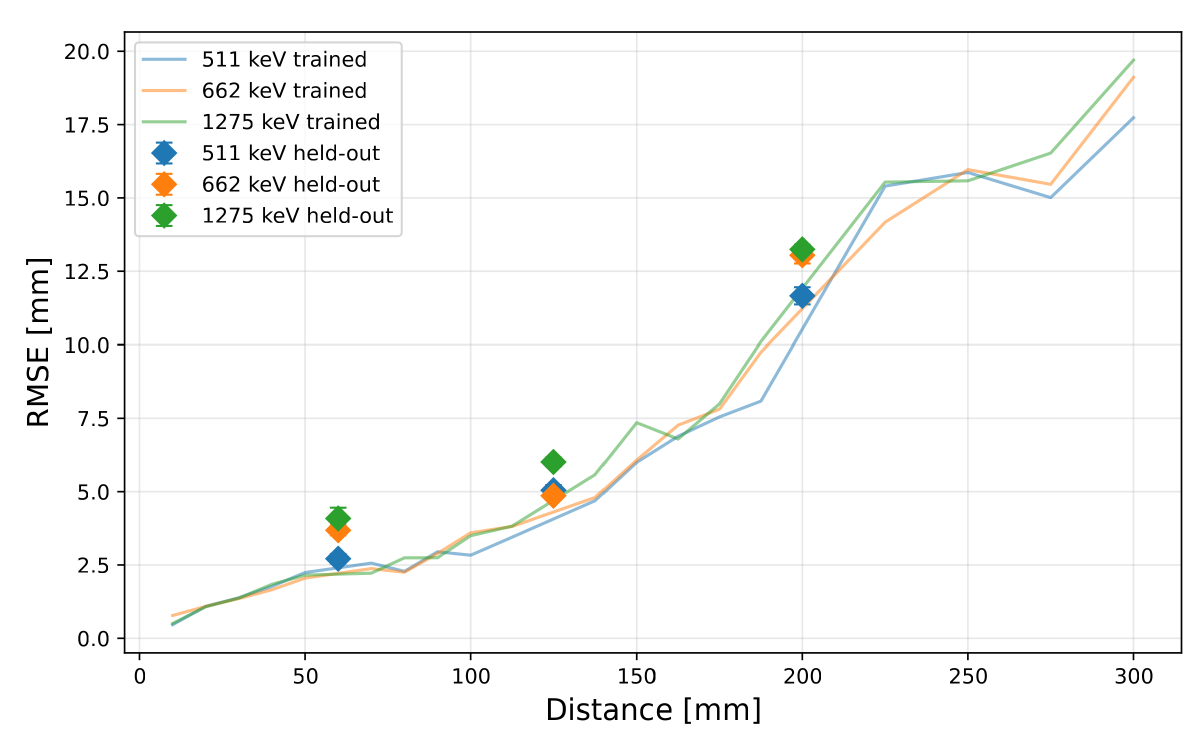}
    \caption{}
    \label{fig:rmse_faces}
  \end{subfigure}
  \begin{subfigure}[b]{0.45\textwidth}
    \centering
    \includegraphics[width=\textwidth]{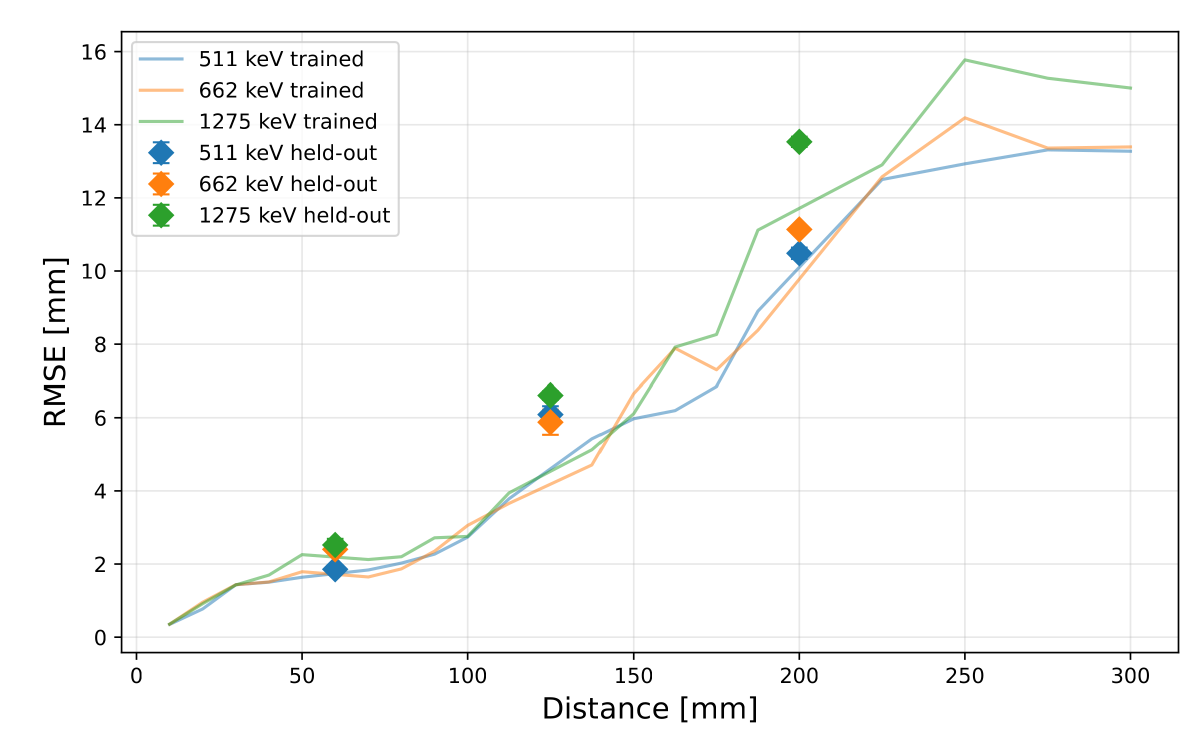}
    \caption{}
    \label{fig:rmse_verts}
  \end{subfigure}
  \caption{The distance interpolation MAEs and RMSEs averaged over 10 random seeds at 511, 662, and 1275 keV energies for faces ((\protect\subref{fig:mae_faces}) and (\protect\subref{fig:rmse_faces})) and vertices ((\protect\subref{fig:mae_verts}) and (\protect\subref{fig:rmse_verts})) of the proposed detector, compared to models' errors trained with 60, 125 and 200 mm distances held-out.}
  \label{fig:inter_plots}
\end{figure}

The validation of the models' predictive performance was performed by testing their interpolation capabilities. The source-to-detector distances at 60, 125, and 200 mm were held out during the model's training for each of the energies. The results of the interpolations can be seen in Table \ref{tab:interpolation}, averaged over 10 random seeds. The MAEs and RMSEs in the Table represent their averaged values for the models at the selected distances, while bias is the measure of the result's average offset from the true distance. The MAEs and RMSEs show the same tendencies as the complete model, increasing with source-to-detector distance and energy of the point sources, and generally lower than half of the respective grid spacings. The MAEs and RMSEs of the interpolated values presented in Fig. \ref{fig:inter_plots} are slightly higher than their trained models' counterparts. MAEs for interpolated distances for both faces (Fig. \ref{fig:mae_faces}) and vertices (Fig. \ref{fig:mae_verts}) are lower than half of the grid spacing at respective distances for every energy considered: less than 4 mm for interpolated distance of 60 mm from the faces and 2 mm for vertices, less than 6 mm for 125 mm, and less than 11 mm at 200 mm for all models. RMSEs show the same trend, with error values for interpolated distance of 60 mm of less than 5 mm for faces (Fig. \ref{fig:rmse_faces}), and less than 3 mm for vertices (Fig. \ref{fig:rmse_verts}). The mid-distance interpolation of 125 mm has RMSEs of $\sim$6.5 mm for both faces and vertices. For the interpolated distance at 200 mm from faces RMSE is $\sim$13.5 mm, and $\sim$11 mm for vertices for energies of 511 and 662 keV, while for 1275 keV the error value is $\sim$13 mm. The bias in Table \ref{tab:interpolation} shows that, overall, predictions are almost perfect, with the maximum offset of 3.5$\pm$0.2 mm for interpolated distance at 125 mm for the model with the point source of 1275 keV energy.

\begin{table}
\centering
\caption{XGBoost regression model results on interpolated distances over 10 random seeds.}
\label{tab:interpolation}
\begin{tabular}{|c|c|c|c|c|}
\hline
    \multicolumn{5}{|c|}{} \\
    \multicolumn{5}{|c|}{FACES} \\
    \multicolumn{5}{|c|}{} \\\hline\hline

    Energy [keV] & True distance [mm] & MAE [mm] & RMSE [mm] &  Bias [mm] \\ \hline

   & 60          & 2.1$\pm$0.1    &    2.7$\pm$0.2    &  1.1$\pm$0.2 \\
511 &    125          & 3.9$\pm$0.2      &    5.0$\pm$0.2    &  0.7$\pm$0.3  \\
   & 200          & 9.5$\pm$0.4      &    11.7$\pm$0.3    &  -1.1$\pm$0.3 \\
    \hline\hline

  &  60          & 3.0$\pm$0.2    &    3.7$\pm$0.2    &  2.5$\pm$0.2 \\
662  &  125          & 3.9$\pm$0.1      &    4.9$\pm$0.1    &  0.4$\pm$0.3  \\
  &  200          & 10.6$\pm$0.3      &    13.0$\pm$0.3    &  -0.2$\pm$0.3 \\
    \hline\hline

   & 60          & 3.5$\pm$0.3    &    4.1$\pm$0.4    &  3.1$\pm$0.3 \\
 1275  & 125          & 5.2$\pm$0.2      &    6.0$\pm$0.1    &  -0.1$\pm$0.3  \\
   & 200          & 10.7$\pm$0.2      &    13.2$\pm$0.2    &  -0.4$\pm$0.2 \\
    \hline\hline
    \multicolumn{5}{|c|}{} \\
    \multicolumn{5}{|c|}{VERTICES} \\
    \multicolumn{5}{|c|}{} \\\hline\hline

    Energy [keV] & True distance [mm] & MAE [mm] & RMSE [mm] &  Bias [mm] \\ \hline
   & 60          & 1.5$\pm$0.1    &    1.9$\pm$0.1    &  0.3$\pm$0.4 \\
511   & 125          & 4.9$\pm$0.2      &    6.1$\pm$0.2    &  2.7$\pm$0.2  \\
   & 200          & 8.7$\pm$0.1      &    10.5$\pm$0.2    &  0.7$\pm$0.1 \\
    \hline\hline

   & 60          & 1.9$\pm$0.1    &    2.4$\pm$0.1    &  -0.3$\pm$0.2 \\
 662  & 125          & 4.9$\pm$0.4      &    5.9$\pm$0.3    &  3.3$\pm$0.4  \\
   & 200          & 9.0$\pm$0.1      &    11.1$\pm$0.1    &  0.9$\pm$0.2 \\
    \hline\hline

   & 60          & 2.0$\pm$0.1    &    2.5$\pm$0.2    &  -0.4$\pm$0.2 \\
 1275  & 125          & 5.7$\pm$0.1      &    6.6$\pm$0.1    &  3.5$\pm$0.2  \\
   & 200          & 10.9$\pm$0.2      &    13.5$\pm$0.1    &  3.1$\pm$0.3 \\
    \hline
\end{tabular}
\end{table}

\section{Discussion}
\label{sec:discussion}
Geant4 simulations were conducted for exploration of the detection sensitivity of a novel radiation detector, capable of determining the position of the source in 4$\pi$. As its readout is positioned on two sensitive volume faces, it raised concerns about the detector's capability for effective collection of gamma rays from all sides. The results have shown that the influence of the geometrical non-uniformity of the design imposes no restrictions on the detection sensitivity from all sides of the detector, as the differences in the efficiencies between the different faces and vertices are less than 1\%. The intrinsic efficiency will suffer from edge effects when measured by a point source. This effect can be seen in Figure \ref{fig:distance_plots}, when the source is positioned in front of the faces, as the intrinsic efficiency of the vertices is higher than for faces at the lowest distance at respective energies. The intrinsic efficiency of the faces increases with the distance of the source, as expected for the larger cross-section, while it decreases for the vertices. The geometrical efficiency falls with increasing distance as well, at the same rate for all energies and all source positions. The highest absolute efficiency is achieved for the lowest source-to-detector distance of 10 mm at 511 keV, but the differences between its values at different energies fade with increasing distance from 100 mm to 300 mm. Due to the detector's small size, with 2 cm$^3$ of sensitive volume, the absolute efficiency of the detector setup converges to <0.1\% for all three tested energies after 100 mm distance. 

The localization of the source prior to the image reconstruction was also one of the goals of the study. The ML regression models were built with the XGBoost library, utilizing the Geant4 simulated data for training, validation, and testing. The results indicate that this algorithm can determine the source-to-detector distance with accuracy higher than half of the training grid spacing for 511, 662, and 1275 keV point source energies, without overfitting and with stable convergence of the models. To test the validity of the predictions of the models, the three distances at 60, 125, and 200 mm were held out during training. Interpolation of these distances has shown that the prediction offset from their true value is less than 3\% in the least accurate model for 1275 keV and 125 mm interpolation distance, while the other models had better predictions. The interpolation results demonstrate that the regression models correctly capture continuous distance dependence, and that the XGBoost algorithm is applicable to the problem of source-to-detector distance determination. As the models' interpolation RMSE remains under 7\% for 200 mm for the 1275 keV source energy, it can be concluded that this approach could be expanded with the training grid of sources set to distances larger than 300 mm. However, the exact constraints of the method require further investigation. The light-sharing principle of the detector can be extended to larger setups, either by increasing the size of the voxels or their number. The concept of source positioning with the XGBoost algorithm could then be applied for training ML models for sources even further away from the detector. 

The detection efficiency and modeling errors could be influenced by the limitations of the simulation setup. The voxels in the detector matrix could have different experimental energy resolutions, depending on their positions. Due to the bright nature of GAGG crystals, with light output of 54 000 photons/MeV, high self-transparency, and compactness of the detector's body, these differences should be small. The experimental construction of the detector would also include plastic black encapsulation for noise suppression, which could contribute to scatter detection. Due to the detector's overall small size and low weight, it would allow for easy portability, as it could be hand-held, mounted on a drone, or on a robotic arm. The absolute position of the source could be determined with a denser grid of point sources surrounding the detector, but the information about the general direction of the gamma ray source and its distance from the detector's face or vertex is enough to speed up image reconstruction by providing initial spatial parameters.

\section{Conclusion}
\label{sec:conclusion}
In this work, we presented the design of a novel omnidirectional segmented radiation detector, utilizing dual perpendicular SiPM readout. Its performance was tested with Monte Carlo simulations in Geant4. We found that detection efficiency for energies relevant to experimental measurements of Na-22 and Cs-137 point sources is not affected by its geometrical non-uniformity. The XGBoost machine learning algorithm successfully predicted the separation distance from the point sources to the detector's faces and vertices. This demonstrates the detector's adequacy for omnidirectional radiation detection, combined with the detector's capability to determine the position of the source through attenuation of the gamma rays and occupancy of the crystals they interact with. The determination of the source's position from raw measurements of the gamma rays could significantly reduce image reconstruction duration and computational requirements when the detector is applied as a Compton camera.

\pagebreak
\section*{Acknowledgments}

This research was funded by the European Union – Next Generation EU and is part of the project Voxel Matrix Detector for Full Solid Angle Radiological Imaging – VMDScan - NPOO.C3.2.R2-I1.06.0084.

\section*{Declaration of AI-assisted tools}

The authors used Claude Opus 4.8 (Anthropic, 2026) to draft Python scripts for data visualization, as well as scripts for the implementation of the XGBoost algorithms. The authors inspected, tested, and revised the scripts before the implementation. Claude did not select the research question, collect data, determine the conclusions, or approve the final manuscript.

\section*{Declaration of data availability}

The data presented in this study will be available upon reasonable request.

\section*{Conflicts of Interest}

The authors declare no conflict of interest.

\pagebreak
\bibliographystyle{unsrt}                                                                          
\bibliography{bibliography}

@article{sato,
	Author = {Sato, Y. and Tanifuji, Y. and Terasaka, Y. and Usami, H. and Kaburagi, M. and Kawabata, K. and Utsugi W. and Kikuchi H. and Takahira S. and Torii, T. },
	Journal = {J. Nucl. Sci. Technol.},
	Number = {9},
	Pages = {965--970},
	Title = {Radiation imaging using a compact Compton camera inside the Fukushima Daiichi Nuclear Power Station building},
	Volume = {55},
	Year = {2018}}

@article{kishimoto,
	Author = {A. Kishimoto and J. Kataoka and T. Kato and T. Miura and T. Nakamori and K. Kamada and S. Nakamura and K. Sato and Y. Ishikawa and K. Yamamura and N. Kawabata and S. Yamamoto},
	Journal = {IEEE Trans. Nucl. Sci.},
	Number = {1},
	Pages = {38--43},
	Title = {Development of a Dual-Sided Readout DOI-PET Module Using Large-Area Monolithic MPPC-Arrays},
	Volume = {60},
	Year = {2013}}

@article{lee3,
	Author = {H. Lee and T. Lee and W. Lee},
	Journal = {IEEE Trans. Nucl. Sci.},
	Number = {12},
	Pages = {226--239},
	Title = {Development of a position-sensitive 4$\pi$ Compton camera based on a single segmented scintillator},
	Volume = {67},
	Year = {2020}}

@article{peng,
	Author = {P. Peng and M. S. Judenhofer and S. R. Cherry},
	Journal = {Phys. Med. Biol.},
	Number = {10},
	Pages = {10LT01},
	Title = {Compton PET: a layered structure PET detector with high performance},
	Volume = {64},
	Year = {2019}}

@article{yamaya,
	Author = {T. Yamaya and T. Mitsuhashi and T. Matsumoto and N. Inadama and F. Nishikido and E. Yoshida and H. Murayama and H. Kawai and M. Suga and M. Watanabe},
	Journal = {Phys. Med. Biol.},
	Number = {21},
	Pages = {6793--6807},
	Title = {A SiPM-based isotropic-3D PET detector X'tal cube with a three-dimensional array of 1 mm3 crystals},
	Volume = {56},
	Year = {2019}}

@article{ghelman,
	Author = {M. Ghelman and N. Kopeika and S. Rotman and N. Ben David and E. Vax and A. Osovizky},
	Journal = {IEEE Trans. Nucl. Sci.},
	Number = {5},
	Pages = {1072--1083},
	Title = {Wide Energetic Response of 4$\pi$ Directional Gamma Detector Based on Combination of Compton Scattering and Photoelectric Effect},
	Volume = {71},
	Year = {2024}}

@article{AGOSTINELLI2003250,
title = {Geant4—a simulation toolkit},
journal = {Nucl. Instrum. Methods Phys. Res. A},
volume = {506},
number = {3},
pages = {250-303},
year = {2003},
author = {S. Agostinelli and J. Allison and K. Amako and J. Apostolakis and H. Araujo and P. Arce and M. Asai and D. Axen and S. Banerjee and G. Barrand and F. Behner and L. Bellagamba and J. Boudreau and L. Broglia and A. Brunengo and H. Burkhardt and S. Chauvie and J. Chuma and R. Chytracek and G. Cooperman and G. Cosmo and P. Degtyarenko and A. Dell'Acqua and G. Depaola and D. Dietrich and R. Enami and A. Feliciello and C. Ferguson and H. Fesefeldt and G. Folger and F. Foppiano and A. Forti and S. Garelli and S. Giani and R. Giannitrapani and D. Gibin and J.J. {Gómez Cadenas} and I. González and G. {Gracia Abril} and G. Greeniaus and W. Greiner and V. Grichine and A. Grossheim and S. Guatelli and P. Gumplinger and R. Hamatsu and K. Hashimoto and H. Hasui and A. Heikkinen and A. Howard and V. Ivanchenko and A. Johnson and F.W. Jones and J. Kallenbach and N. Kanaya and M. Kawabata and Y. Kawabata and M. Kawaguti and S. Kelner and P. Kent and A. Kimura and T. Kodama and R. Kokoulin and M. Kossov and H. Kurashige and E. Lamanna and T. Lampén and V. Lara and V. Lefebure and F. Lei and M. Liendl and W. Lockman and F. Longo and S. Magni and M. Maire and E. Medernach and K. Minamimoto and P. {Mora de Freitas} and Y. Morita and K. Murakami and M. Nagamatu and R. Nartallo and P. Nieminen and T. Nishimura and K. Ohtsubo and M. Okamura and S. O'Neale and Y. Oohata and K. Paech and J. Perl and A. Pfeiffer and M.G. Pia and F. Ranjard and A. Rybin and S. Sadilov and E. {Di Salvo} and G. Santin and T. Sasaki and N. Savvas and Y. Sawada and S. Scherer and S. Sei and V. Sirotenko and D. Smith and N. Starkov and H. Stoecker and J. Sulkimo and M. Takahata and S. Tanaka and E. Tcherniaev and E. {Safai Tehrani} and M. Tropeano and P. Truscott and H. Uno and L. Urban and P. Urban and M. Verderi and A. Walkden and W. Wander and H. Weber and J.P. Wellisch and T. Wenaus and D.C. Williams and D. Wright and T. Yamada and H. Yoshida and D. Zschiesche}
}

@article{ref-parallel,
      author        = {{O. Tang}},
    title = "{GNU Parallel 20210822 ('Kabul')}",
    journal = "Zenodo",
    year = "2022"
}

@article{root,
title = {ROOT — An object oriented data analysis framework},
journal = {Nucl. Instrum. Methods Phys. Res. A},
volume = {389},
number = {1},
pages = {81-86},
year = {1997},
author = {Rene Brun and Fons Rademakers},
}

@conference{hndpaper,
  	author       = {{A.M. Kožuljević} and Jazvac, G. and Lotina, L. and Pavelić, L.},
  	title        = {Segmented Compton Camera with Dual Perpendicular Silicon Photomultiplier Read-Out for 4$\pi$ Gamma Ray Imaging},
  	booktitle    = {Proceedings of the 15th International Conference of the Croatian Nuclear Society},
  	month        = {},
  	year         = {2026},
  	address      = {Zadar, Croatia, May 31--June 3},
  	organization    = {Croatian Nuclear Society}}

@article{xai,
    author = {Bandstra, Mark S. AND Curtis, Joseph C. AND Ghawaly, Jr, James M. AND Jones, A. Chandler AND Joshi, Tenzing H. Y.},
    journal = {PLOS ONE},
    title = {Explaining machine-learning models for gamma-ray detection and identification},
    year = {2023},
    month = {06},
    volume = {18},
    pages = {1-21},
    number = {6},

}

@inproceedings{xgboost,
author = {Chen, Tianqi and Guestrin, Carlos},
title = {XGBoost: A Scalable Tree Boosting System},
year = {2016},
address = {New York, NY, USA},
booktitle = {Proceedings of the 22nd ACM SIGKDD International Conference on Knowledge Discovery and Data Mining},
pages = {785–794},
numpages = {10},
location = {San Francisco, California, USA},
series = {KDD '16}
}

@article{hpge_acc,
title = {Accuracy of machine learning algorithms for HPGe detector efficiency determination},
journal = {Radiation Physics and Chemistry},
volume = {239},
pages = {113328},
year = {2026},
author = {B. González-González and U. Abascal-Ruiz and M. Villa-Alfageme and S. Hurtado-Bermúdez}
}

@article{scikit-learn,
  title={Scikit-learn: Machine Learning in {P}ython},
  author={Pedregosa, F. and Varoquaux, G. and Gramfort, A. and Michel, V.
          and Thirion, B. and Grisel, O. and Blondel, M. and Prettenhofer, P.
          and Weiss, R. and Dubourg, V. and Vanderplas, J. and Passos, A. and
          Cournapeau, D. and Brucher, M. and Perrot, M. and Duchesnay, E.},
  journal={Journal of Machine Learning Research},
  volume={12},
  pages={2825--2830},
  year={2011}
}

@INPROCEEDINGS{uher,
  author={Uher, J. and Frojdh, C. and Jakubek, J. and Pospisil, S. and Thungstrom, G. and Vykydal, Z.},
  booktitle={2007 IEEE Nuclear Science Symposium Conference Record}, 
  title={Directional radiation detector}, 
  year={2007},
  volume={2},
  number={},
  pages={1162-1165},
 }

@article{Makek2020_crystals,
   author = {Makek, Mihael and Bosnar, Damir and Ko\v{z}uljevi\'{c}, Ana Marija and Paveli\'{c}, Luka},
   title = {{Investigation of GaGG:Ce with TOFPET2 ASIC Readout for Applications in Gamma Imaging Systems}},
   journal = {{Crystals}},
   volume = {10},
   pages = {1073},
   year = {2020},
   doi = {https://doi.org/10.3390/cryst10121073}
}

@article{GHELMAN2025170246,
title = {4$\pi$ directional radiation detector based on Compton scattering with channel reduction method},
journal = {Nuclear Instruments and Methods in Physics Research Section A: Accelerators, Spectrometers, Detectors and Associated Equipment},
volume = {1072},
pages = {170246},
year = {2025},
author = {Max Ghelman and Natan Kopeika and Stanley Rotman and Zohar Davidov and Alon Osovizky},
}

@article{DASH2023167808,
title = {Design and characteristics of a novel single-plane Compton gamma camera based on GAGG scintillators readout by SiPMs},
journal = {Nuclear Instruments and Methods in Physics Research Section A: Accelerators, Spectrometers, Detectors and Associated Equipment},
volume = {1047},
pages = {167808},
year = {2023},
author = {Om Prakash Dash and Tomislav Bokulić and Damir Bosnar and Petar Žugec and Mihael Makek},
}

@article{RAWOOLSULLIVAN20081986,
title = {A simple algorithm for estimation of source-to-detector distance in Compton imaging},
journal = {Applied Radiation and Isotopes},
volume = {66},
number = {12},
pages = {1986-1991},
year = {2008},
author = {Mohini W. Rawool-Sullivan and John P. Sullivan and Shawn R. Tornga and Steven P. Brumby},

}

@article{llosa,
title = {Improving Compton camera imaging of multi-energy radioactive sources by using machine learning algorithms for event selection},
journal = {Radiation Physics and Chemistry},
volume = {226},
pages = {112166},
year = {2025},
author = {J. Pérez-Curbelo and J. Roser and E. Muñoz and L. Barrientos and V. Sanz and G. Llosá},

}

@article{ml-nearfield1,
title = {Localization of radioactive source using a network of small form factor CZT sensors},
journal = {Journal of Instrumentation},
volume = {15},
pages = {C04015},
year = {2020},
author = {G. Fragkos and K. Karafasoulis and A. Kyriakis and C. Potiriadis},

}

@article{ml-nearfield,
title = {Simulated near-field radioactive source localization in 3D with Coded Aperture and Convolutional Neural Networks},
journal = {Journal of Instrumentation},
volume = {20},
pages = {C01046},
year = {2025},
author = {K. Karafasoulis and I. Kaissas and C. Papadimitropoulos and K. Potiriadis and C.P. Lambropoulos},

}

@article{DURBIN2022166905,
title = {Experimental tests of Gamma-ray Localization Aided with Machine-learning (GLAM) capabilities},
journal = {Nuclear Instruments and Methods in Physics Research Section A: Accelerators, Spectrometers, Detectors and Associated Equipment},
volume = {1038},
pages = {166905},
year = {2022},
author = {Matthew Durbin and Ryan Sheatsley and Patrick McDaniel and Azaree Lintereur},
}

@Article{sspd-ml,
AUTHOR = {Simhony, Yoav and Segal, Alex and Amrani, Ofer and Etzion, Erez},
TITLE = {From Light to Energy: Machine Learning Algorithms for Position and Energy Deposition Estimation in Scintillator–SiPM Detectors},
JOURNAL = {Sensors},
VOLUME = {26},
YEAR = {2026},
NUMBER = {1},
ARTICLE-NUMBER = {101},
}

\end{document}